\documentclass[aps,prd,preprintnumbers,groupedaddress,nofootinbib,amssymb,eqsecnum,notitlepage]{revtex4-1}

\usepackage[dvipdfmx]{graphicx}
\usepackage{bm}
\usepackage{amsmath}
\usepackage{color}
\usepackage{amsfonts}
\usepackage{here}
\usepackage{graphicx}
\usepackage{amsmath,amsthm,amssymb}
\usepackage{bm}
\allowdisplaybreaks[1]

\usepackage{amsfonts}
\usepackage{dcolumn}
\usepackage{cleveref}
\begin{document}

\newcommand{\newc}{\newcommand}
\newc{\tif}{\tilde{f}}
\newc{\tih}{\tilde{h}}
\newc{\tip}{\tilde{\phi}}
\newc{\tiA}{\tilde{A}}

\newcommand{\rk}[1]{{\color{red} #1}}
\newcommand{\mr}[1]{\mathrm{#1}}
\newcommand{\mL}[1]{\mathcal{#1}}
\newcommand{\vc}[1]{\boldsymbol{#1}}
\renewcommand{\arraystretch}{1.7}
\newcommand{\ben}{\begin{eqnarray}}
\newcommand{\een}{\end{eqnarray}}
\newc{\be}{\begin{equation}}
\newc{\ee}{\end{equation}}
\newc{\ba}{\begin{eqnarray}}
\newc{\ea}{\end{eqnarray}}
\newc{\bea}{\begin{eqnarray*}}
\newc{\eea}{\end{eqnarray*}}
\newc{\D}{\partial}
\newc{\ie}{{\it i.e.} }
\newc{\eg}{{\it e.g.} }
\newc{\etc}{{\it etc.} }
\newc{\etal}{{\it et al.}}
\newcommand{\nn}{\nonumber}
\newc{\ra}{\rightarrow}
\newc{\lra}{\leftrightarrow}
\newc{\lsim}{\buildrel{<}\over{\sim}}
\newc{\gsim}{\buildrel{>}\over{\sim}}
\newc{\aP}{\alpha_{\rm P}}
\newc{\dphi}{\delta\phi}
\newc{\da}{\delta A}
\newc{\tp}{\dot{\phi}}
\newc{\rd}{{\rm d}}
\newc{\rE}{{\rm E}}
\newc{\re}{r_{\rm E}}
\newc{\Mpl}{M_{\rm pl}}

\title{Neutron stars in $f(R)$ gravity and scalar-tensor theories}

\author{Ryotaro Kase and Shinji Tsujikawa}

\affiliation{Department of Physics, Faculty of Science, 
Tokyo University of Science, 1-3, Kagurazaka,
Shinjuku-ku, Tokyo 162-8601, Japan}

\date{\today}

\begin{abstract}

In $f(R)$ gravity and Brans-Dicke theory with scalar 
potentials, we study the structure of neutron stars  
on a spherically symmetric and static background for two equations 
of state: SLy and FPS. 
In massless BD theory, the presence of a scalar coupling $Q$ 
with matter works to change the star radius in comparison to 
General Relativity, while the maximum allowed mass of neutron stars 
is hardly modified for both SLy and FPS equations of state. 
In Brans-Dicke theory with the massive potential $V(\phi)=m^2 \phi^2/2$, 
where $m^2$ is a positive constant, we show the difficulty of 
realizing neutron star solutions with a stable field profile due to the 
existence of an exponentially growing mode outside the star. 
As in $f(R)$ gravity with the $R^2$ term, this property
is related to the requirement of extra boundary 
conditions of the field at the surface of star. 
For the self-coupling potential $V(\phi)=\lambda \phi^4/4$, 
this problem can be circumvented by the fact that the second 
derivative $V_{,\phi \phi}=3\lambda\phi^2$ 
approaches 0 at spatial infinity.
In this case, we numerically show the existence of neutron star 
solutions for both SLy and FPS equations of state 
and discuss how the mass-radius relation 
is modified as compared to General Relativity.

\end{abstract}


\maketitle

%%%%%%%%%%%%%%%%%%%%%%%%%%%%%%
\section{Introduction}
\label{Introsec}
%%%%%%%%%%%%%%%%%%%%%%%%%%%%%%

The dawn of gravitational-wave (GW) astronomy opened up a new 
possibility for probing the physics in the strong-field 
regime \cite{Abbott:2016blz}. 
The accuracy of General Relativity (GR) is well confirmed 
on the weak gravitational background \cite{Will:2014kxa}, 
but the theory can be subject to modifications in the region of high density 
with the large scalar curvature $R$. 
Now, we are entering the golden era in which the deviation from 
GR can be tested from the GW observations of strong gravitational 
sources such as black holes (BHs) \cite{TheLIGOScientific:2016src} 
and neutron stars (NSs) \cite{TheLIGOScientific:2017qsa}.

One of the simplest modifications from GR is known as $f(R)$  
theories, in which the Lagrangian contains nonlinear functions 
of $R$ \cite{Bergmann,Ruz}. 
In the Starobinsky $f(R)$ model characterized by the 
Lagrangian $f(R)=R+R^2/(6m^2)$ \cite{Staro}, 
where $m^2$ is a positive mass squared, the existence of 
the $R^2$ term can drive cosmic inflation in the early Universe. 
In the same manner, $f(R)$ theories have been extensively 
applied to the physics of late-time 
cosmic acceleration \cite{fRearly1,fRearly1b,fRearly2,fRearly3,fR1,fR2,fR3,fR4,fR5}. 
The existence of higher-curvature term in $f(R)$ theories can also be 
potentially important in local objects on the strong gravitational background. 
In this vein, many papers devoted to the study of spherically symmetric 
and static BH \cite{delaCruzDombriz:2009et,Nelson:2010ig,Kehagias:2015ata,Canate:2015dda,Yu:2017uyd,Canate:2017bao,Sultana:2018fkw} and NS \cite{Cooney:2009rr,Arapoglu:2010rz,Orellana:2013gn,Astashenok:2013vza,Ganguly:2013taa,Yazadjiev,Capozziello:2015yza,Resco:2016upv,Feng:2017hje} 
solutions in the Starobinsky model and pure $R^2$ gravity.
We note that $f(R)$ theories are equivalent to Brans-Dicke (BD) 
theories \cite{Brans} with a scalar potential of the gravitational 
origin \cite{Ohanlon,Chiba03} (see also Ref.~\cite{Chakraborty}). 
In BD theories with the potential of a positive mass squared, 
there is the ``no-hair'' theorem of BHs forbidding the 
existence of a nontrivial scalar hair \cite{Hawking:1972qk,Bekenstein:1995un,Sotiriou:2011dz}. 
This does not allow the presence of hairy BH solutions 
in the Starobinsky model \cite{delaCruzDombriz:2009et,Nelson:2010ig,Yu:2017uyd,Canate:2017bao}.

In BD theories, the nonminimal coupling to gravity mediates the fifth force 
between the scalar field $\phi$ and matter \cite{Fujii}. 
This property is particularly transparent in the Einstein frame where 
the field $\phi$ directly interacts with matter with a universal 
coupling constant $Q$ \cite{chame1,chame2,Amendola:2006kh,Yoko}. 
For example, the $f(R)$ gravity in the metric formalism 
corresponds to $Q=-1/\sqrt{6}$ \cite{fRreview}. 
This scalar-matter coupling plays an important role for 
studying the existence of NS solutions in BD theories and $f(R)$ gravity. 
In the Starobinsky $f(R)$ model mentioned above, the potential in the Einstein frame is given 
by $V_{\rE}(\phi)=(3/4)m^2M_{\rm pl}^2 [1-e^{-\sqrt{6}\phi/(3M_{\rm pl})}]^2$, 
where $\phi=(\sqrt{6}M_{\rm pl}/2) \ln [1+R/(3m^2)]$ and $M_{\rm pl}$ 
is the reduced Planck mass \cite{fRreview}. 
In the regime $|\phi| \ll M_{\rm pl}$, this potential 
reduces to $V_{\rE}(\phi) \simeq m^2 \phi^2/2$, so the scalar field has a constant 
mass $m$ with the matter coupling $Q=-1/\sqrt{6}$.
 
If we apply $f(R)$ theories to NSs on the spherically symmetric and static 
background, both the potential and matter coupling contribute to 
the scalar-field equation inside the star. 
For the vacuum exterior, the $\phi$ derivative of Einstein-frame 
potential $V_{\rE}(\phi)$ mostly determines the field profile 
outside the star. For the massive potential 
$V_{\rE}(\phi)=m^2 \phi^2/2$, the field equation contains 
a growing-mode solution of the form $\phi \propto e^{mr}/r$ 
outside the body, where $r$ is the distance from the center of 
symmetry. In this case, we do not realize the asymptotic flat 
boundary conditions $\phi \to 0$ and $\rd \phi/\rd r \to 0$  
at spatial infinity. This exponential growth of $\phi$ can be avoided 
by imposing the boundary conditions $\phi=0$ and $\rd \phi/\rd r=0$ 
at the {\it surface} of star, which amounts to considering the Schwarzschild 
exterior without any scalar-field contribution to the metric. 
However, as claimed in Ref.~\cite{Ganguly:2013taa}, the existence of 
such additional conditions does not allow the natural realization of 
NS solutions for arbitrary equations of state (EOSs). 
This property holds not only for $f(R)$ gravity with the massive potential 
$V_{\rE}(\phi)=m^2 \phi^2/2$ but also for the Starobinsky $f(R)$ model.

On the other hand, the $f(R)$ models of late-time cosmic acceleration \cite{fR1,fR2,fR3,fR4} are 
constructed to have a density-dependent effective scalar mass $m_{\phi}$ to accommodate 
the chameleon mechanism \cite{chame1,chame2,Faulkner:2006ub,Capozziello:2007eu} 
in over-density regions with the nonrelativistic 
background (see also Ref.~\cite{Brax}). 
In such cases the scalar field is heavy inside the NS, but 
it can be practically massless outside the star. 
In spite of accessible curvature singularities in those $f(R)$ 
dark energy models \cite{fR2,Frolov:2008uf,Kobayashi:2008tq,Kobayashi:2008wc}, 
the existence of relativistic stars was shown for a constant density profile \cite{Upadhye:2009kt} 
and for a polytropic EOS \cite{Babichev:2009td,Babichev:2009fi}.
In BD theories with the inverse power-law potential 
$V_{\rE}=M^{4+n}\phi^{-n}$ ($n>0$), there are also chameleon-like solutions 
for relativistic stars with the constant density \cite{Tsujikawa:2009yf}. 
As in the standard chameleon solution on the Minkowski 
background \cite{chame1,chame2}, 
the field $\phi$ and its $r$ derivative do not vanish outside the star.
Hence the boundary conditions $\phi=0$ and $\rd \phi/\rd r=0$ 
at the surface of star ($r=r_s$) are not mandatory for the nearly massless 
scalar field at the distance $r>r_s$.

In this paper, we study the NS solutions in BD theories with the scalar potential
and the general coupling $Q$ for two realistic EOSs: SLy \cite{SLY} 
and FPS \cite{FPS}. For this purpose, we use the analytic representations 
of these two EOSs presented in Ref.~\cite{Haensel:2004nu}. 
Our analysis is sufficiently general in that it covers massless BD theories 
and $f(R)$ gravity as special cases. In particular, we would like to clarify the 
difference of NS solutions between the potential $V(\phi)$ 
with a constant mass and the potential allowing the asymptotic 
behavior $V_{,\phi\phi} \equiv {\rm d^2}V/{\rm d}\phi^2 \to 0$ at spatial infinity.
For this purpose, we consider BD theories with the self-coupling 
potential $V(\phi)=\lambda \phi^4/4$ in the Jordan frame.
The same potential was also introduced in the Einstein frame 
for accommodating the chameleon mechanism on the nonrelativistic background \cite{Gubser:2004uf}. 
It is not yet clear whether the similar chameleon-like solutions arise 
on the relativistic background with realistic EOSs. 
We show the existence of new NS solutions for both SLy and FPS 
EOSs, despite a different property of the field profile from that 
on the weak gravitational background.

The NS solution with the potential $V(\phi)=\lambda \phi^4/4$ is in contrast to that in 
massive BD theories with $V(\phi)=m^2 \phi^2/2$. 
While the scalar field in the latter case is subject to exponential growth by 
the constant mass $m$ outside the body, the former potential evades this 
problem due to the property that the mass squared  
$V_{,\phi\phi}$ approaches 0 as $r \to \infty$. 
We compute the mass $M$ and radius $r_s$ of NSs in BD theories 
with/without the potential $V(\phi)=\lambda \phi^4/4$ 
in order to see the signature for the modification of gravity from GR.
For the purely massless case ($\lambda=0$), the modification to the 
radius $r_s$ tends to be significant for increasing $|Q|$ of order 0.1, 
while the maximum NS mass is hardly changed.
As $\lambda$ increases, the theoretical curve of the mass-radius relation 
approaches that in GR by reflecting the fact that the field tends to be 
heavy inside the star.

This paper is organized as follows.
In Sec.~\ref{backsec}, we derive the full equations of motion on the spherically 
symmetric and static background in BD theories with the scalar potential $V(\phi)$ 
in both Jordan and Einstein frames. 
We also discuss the boundary conditions at the center of 
star and at spatial infinity. 
In Sec.~\ref{masslesssec}, we study the mass-radius relation of NSs 
in BD theories with $V(\phi)=0$ and investigate how much 
modification from GR arises for different couplings $Q$.
In Sec.~\ref{massivesec}, we consider the massive potential 
$V(\phi)=m^2 \phi^2/2$ and show the difficulty of obtaining
NS solutions consistent with the boundary conditions 
at spatial inifnity. 
In Sec.~\ref{selfsec}, we investigate how the NS solutions can 
be realized by the self-coupling potential $V(\phi)=\lambda \phi^4/4$ 
and compare the mass-radius relation with that in GR.
Sec.~\ref{consec} is devoted to conclusions.

In this paper we adopt the natural units $c=\hbar=1$, 
where $c$ is the speed of light and $\hbar$ is reduced Planck constant.  
When these fundamental constants are needed in numerical computations, 
we recover them and use their concrete values $c=2.9979\times10^{10}~{\rm cm\cdot s^{-1}}$ 
and $\hbar=1.0546\times10^{-27}~{\rm erg\cdot s}$, together 
with the Newton gravitational constant
$G=6.6743\times10^{-8}~{\rm g^{-1}\cdot cm^3\cdot s^{-2}}$.

%%%%%%%%%%%%%%%%%%%%%%%%%%%%%%
\section{Equations of motion}
\label{backsec}
%%%%%%%%%%%%%%%%%%%%%%%%%%%%%%

We begin with the action of scalar-tensor theories accommodating BD theories 
with a scalar potential $V(\phi)$,
\be
{\cal S}=\int {\rm d}^4x \sqrt{-g} \left[ \frac{M_{\rm pl}^2}{2} F(\phi)R
+ \left( 1-6Q^2 \right)F(\phi) X -V(\phi) \right]
+\int {\rm d}^4 x\,{\cal L}_m \left( g_{\mu \nu}, \Psi_m \right)\,,
\label{actionJ}
\ee
where $g$ is the determinant of metric tensor $g_{\mu \nu}$, 
$F(\phi)$ is a function of a scalar field $\phi$, 
$R$ is the Ricci scalar, $Q$ is a constant, 
$X=-g^{\mu \nu} \partial_{\mu} \phi \partial_{\nu} \phi/2$, 
and ${\cal L}_m$ is the action of matter fields $\Psi_m$. 
In Ref.~\cite{Yoko}, it was shown that BD theories \cite{Brans} 
with the potential 
corresponds to the nonminimal coupling: 
\be
F(\phi)=e^{-2 Q\phi/M_{\rm pl}}\,.
\label{Fphi}
\ee
In the limit that $Q \to 0$, the action (\ref{actionJ}) reduces to that 
of a canonical scalar field. 
The constant $Q$ characterizes the coupling between 
the field $\phi$ and the gravity sector. 
This coupling constant is related to the 
BD parameter $\omega_{\rm BD}$ as 
$2Q^2=1/(3+2\omega_{\rm BD})$ \cite{Yoko}. 
The matter energy-momentum tensor is defined by 
$T_{\mu \nu}=-(2/\sqrt{-g}) \delta {\cal L}_m/ \delta g^{\mu \nu}$.
Assuming that the matter fields are minimally coupled to gravity, 
there is the continuity equation 
\be
\nabla^{\mu} T_{\mu \nu}=0\,,
\label{coneq}
\ee
where $\nabla^{\mu}$ is the covariant derivative operator. 

The metric $f(R)$ gravity given by the action 
\be
{\cal S}_{f(R)}=\int {\rm d}^4x \sqrt{-g}\,
\frac{M_{\rm pl}^2}{2} f(R)
\label{actionfR}
\ee
belongs to a subclass of the graviton-scalar action in Eq.~(\ref{actionJ}), 
with the correspondence \cite{fRreview}
\be
Q=-\frac{1}{\sqrt{6}}\,,\qquad 
V(\phi)=\frac{M_{\rm pl}^2}{2} \left( FR-f \right)\,,\qquad 
F=\frac{\partial f}{\partial R}=e^{-2 Q\phi/M_{\rm pl}}\,.
\label{fRre}
\ee
For $f(R)$ containing nonlinear functions in $R$, the scalar degree of freedom 
$\phi$ arises from the gravity sector. 
In this case the field potential $V(\phi)$ does not vanish, so the 
field $\phi$ generally has a nonvanishing effective mass.

In string theory, the low-energy effective action contains the so-called 
dilaton field $\Phi$ coupled to gravity \cite{Gas1,Gas2}. 
The lowest-order graviton-dilaton action in 4-dimensional 
spacetime takes the form
\be
{\cal S}_{\rm dilaton}=\int {\rm d}^4x \sqrt{-g}\,
\frac{e^{-\Phi/M_{\rm pl}}}{2} \left( M_{\rm pl}^2 R
+g^{\mu \nu} \partial_{\mu} \Phi \partial_{\nu} \Phi
\right)\,.
\label{actiondi}
\ee
After the field redefinition $\Phi \to 2Q \phi$, the action 
(\ref{actiondi}) reduces to the graviton-scalar action 
in Eq.~(\ref{actionJ}), with the correspondences 
$Q^2=1/2$ and $V(\phi)=0$. 
Thus, the massless dilaton can be also accommodated 
in BD theory  with the specific coupling $Q^2=1/2$.

In this paper, we will consider BD theories with general couplings $Q$ 
including metric $f(R)$ theories and dilaton gravity
in the absence/presence of $V(\phi)$.
We deal with the Jordan frame given by the action (\ref{actionJ}) 
as a physical frame and derive the equations of motion on 
the spherically symmetric and static background.
We also discuss the field configuration in the Einstein frame 
in which the matter sector is directly coupled to $\phi$. 

\subsection{Jordan frame}

We study the NS solutions on the spherically symmetric 
and static background given by the line element
\be
{\rm d}s^2=-f(r) \rd t^2+h^{-1}(r) \rd r^2 
+r^2 \left( \rd \theta^2+\sin^2 \theta\,\rd \varphi^2 \right)\,,
\label{metJ}
\ee
where $f$ and $h$ are functions of the distance $r$ 
from the center of symmetry. 
For the matter sector, we consider a perfect fluid whose 
energy-momentum tensor is given by 
$T^{\mu}_{\nu}={\rm diag}\,(-\rho (r), P(r), P(r), P(r))$, 
where $\rho(r)$ is the energy density and 
$P(r)$ is the pressure. 
{}From the continuity equation (\ref{coneq}), we obtain
\be
P'+\frac{f'}{2f} \left( \rho+P \right)=0\,,
\label{macon}
\ee
where a prime represents the derivative with respect to $r$.
To relate $P$ with $\rho$ for realistic NSs, we resort to the 
analytic representations of SLy and FPS EOSs \cite{Haensel:2004nu}. 
Introducing the notations 
\be
\xi=\log_{10} (\rho/{\rm g \cdot cm}^{-3})\,,\qquad 
\zeta=\log_{10} (P/{\rm dyn \cdot cm}^{-2})\,,
\label{xizeta}
\ee
the two EOSs can be parameterized as
\ba
\zeta (\xi) &=&
\frac{a_1+a_2 \xi+a_3 \xi^3}{1+a_4 \xi}
f_{0} \left(a_5(\xi-a_6) \right)+\left( a_7+a_8 \xi \right) 
f_{0} \left(a_9(a_{10}-\xi) \right)
+\left( a_{11}+a_{12} \xi \right) 
f_{0} \left(a_{13}(a_{14}-\xi) \right) \nonumber \\
& &
+\left( a_{15}+a_{16} \xi \right) 
f_{0} \left(a_{17}(a_{18}-\xi) \right)\,,
\label{zeta}
\ea
where 
\be
f_0(x)=\left( e^x+1 \right)^{-1}\,,
\ee
and the coefficients $a_{1, \cdots , 18}$ for the SLy and FPS are given in 
Table 1 of Ref.~\cite{Haensel:2004nu}. 
Taking account of additional functions to Eq.~(\ref{zeta}), 
it is also possible to accommodate other EOSs like 
BSk19, BSk20, and BSk21 \cite{Potekhin:2013qqa}.

Instead of the metric $h$, it is convenient to introduce the 
mass function ${\cal M}(r)$ defined by 
\be
h(r)=1-\frac{2G {\cal M}(r)}{r}\,,
\label{hr}
\ee
where the gravitational constant $G$ is related to $M_{\rm pl}$ 
as $G=(8\pi M_{\rm pl}^2)^{-1}$. 
We define the ADM mass $M$ of the star, as
\be
M \equiv \lim_{r \to \infty} {\cal M}(r)=
\frac{r}{2G} \left( 1-h \right) \biggl|_{r \to \infty}\,.
\ee
The star radius $r_s$ is determined by the condition 
\be
P (r_s)=0\,.
\ee
Varying the action (\ref{actionJ}) with respect to $g_{\mu \nu}$ 
and $\phi$, the equations of motion on the background 
(\ref{metJ}) read
\ba
\frac{f'}{f} &=& -\frac{2M_{\rm pl}^2 (h-1)
-2F^{-1} r^2 (P-V)+hr \phi' [(6Q^2-1)r \phi'-8Q M_{\rm pl}]
}{2h r M_{\rm pl} (M_{\rm pl}-Qr \phi')}\,,\label{eq1}\\
{\cal M}' &=& 4\pi F^{-1} r^2 \left[ (1-2Q^2) \rho+6Q^2 P
+(1-8Q^2)V-2QM_{\rm pl} V_{,\phi} \right] \nonumber \\
& & +\phi' \frac{2QM_{\rm pl}{\cal M}+8Q M_{\rm pl} \pi r^3 F^{-1}
(P-V)+r\phi'(4 \pi r M_{\rm pl}^2-{\cal M})(1+2Q^2)}{2M_{\rm pl}(M_{\rm pl}-Qr \phi')}\,,\label{eq2}\\
\phi'' &=& -\frac{\phi'}{2M_{\rm pl}^2 rh} 
\left[ 2(h+1)M_{\rm pl}^2+r^2 F^{-1} \{ 
P-\rho+2 QM_{\rm pl} V_{,\phi}-2V
+2(\rho-3P+4V)Q^2 \} \right] \nonumber \\
& &+\frac{1}{M_{\rm pl}hF}
\left[ 4QV+V_{,\phi}M_{\rm pl}
+Q(\rho-3P) \right]\,,\label{eq3}
\ea
where $V_{,\phi} \equiv \rd V/\rd \phi$. 
On using these equations, the Ricci scalar is expressed as 
\be
R=\frac{1}{M_{\rm pl}^2} \left\{ (1-6Q^2) \left[ h \phi'^2
+\left( 4V+ \rho-3P \right) 
e^{2Q \phi/M_{\rm pl}} \right]
-6M_{\rm pl}Q V_{,\phi}e^{2Q \phi/M_{\rm pl}} 
\right\}\,,
\label{Rphi}
\ee
which shows that not only the matter density and pressure but also 
the field kinetic energy and potential generally contribute to $R$.

The regularities of solutions at the center of NSs demands
the following boundary conditions 
\be
f' (r=0)=0\,,\qquad h'(r=0)=0\,,\qquad \phi'(r=0)=0\,, 
\qquad \rho'( r=0)=0\,.
\label{boun0}
\ee
As long as the mass function has the dependence ${\cal M}(r) \propto r^3$ 
at leading order, we also have $h(r=0)=1$ from Eq.~(\ref{hr}). 
For the consistency with Eq.~(\ref{boun0}), 
we expand $f, h, \phi, \rho$ around $r=0$, as
\be
f(r)=f_0+\sum_{n=2}^{\infty} f_n r^n\,,\qquad
h(r)=1+\sum_{n=2}^{\infty} h_n r^n\,,\qquad 
\phi(r)=\phi_0+\sum_{n=2}^{\infty} \phi_n r^n\,,\qquad 
\rho(r)=\rho_0+\sum_{n=2}^{\infty} \rho_n r^n\,,
\label{exp}
\ee
where $f_0, f_n, h_n, \phi_0, \phi_n, \rho_0, \rho_n$ are constants. 
On using Eq.~(\ref{zeta}), the pressure can be  
written in the form $P(r)=P_0+\sum_{n=2}^{\infty}P_n r^n$, 
where $P_0, P_n$ are constants.  
We also expand the potential in terms of the Taylor series, as
\be
V(\phi)=V(\phi_0)+\sum_{n=1}^{\infty} 
\frac{1}{n!}
\frac{\rd^n V}{\rd \phi^n}\biggl|_{\phi=\phi_0}
(\phi-\phi_0)^n\,.
\ee
Then, the solutions consistent with Eqs.~(\ref{macon}) and 
(\ref{eq1})-(\ref{eq3}) around $r=0$ are given by 
\ba
f(r) &=& f_0 \left\{ 1+\frac{[(1+2Q^2)\rho_0+3(1-2Q^2)P_0
+2(4Q^2-1)V(\phi_0) +2QM_{\rm pl}V_{,\phi} (\phi_0)]}
{6M_{\rm pl}^2 e^{-2Q \phi_0/M_{\rm pl}}} r^2 +{\cal O} (r^4)
\right\}\,,\label{fexpan}\\
h(r) &=& 1-\frac{(1-2Q^2)\rho_0+6Q^2 P_0
+(1-8Q^2)V(\phi_0) -2QM_{\rm pl}V_{,\phi} (\phi_0)}
{3M_{\rm pl}^2e^{-2Q \phi_0/M_{\rm pl}}} r^2 +{\cal O} (r^4)\,,\\
\phi(r) &=& \phi_0+\frac{Q\{\rho_0-3P_0
+4V(\phi_0) \} +M_{\rm pl}V_{,\phi} (\phi_0)}
{6M_{\rm pl}e^{-2Q \phi_0/M_{\rm pl}}} r^2 +{\cal O} (r^4)\,,
\label{phiexpan}\\
P(r)&=& P_0-\frac{[(1+2Q^2)\rho_0+3(1-2Q^2)P_0
+2(4Q^2-1)V(\phi_0) +2QM_{\rm pl}V_{,\phi} (\phi_0)] (\rho_0+P_0)}
{12M_{\rm pl}^2 e^{-2Q \phi_0/M_{\rm pl}}} r^2 +{\cal O} (r^4)\,.
\label{Pexpan}
\ea
Both the coupling $Q$ and the $\phi$ derivative of $V(\phi)$ 
lead to the variation of $\phi$ around the center of body.
They also give rise to modifications to $f$, $h$, $P$ 
in comparison to the theories with $Q=0$ and $V(\phi)=0$.

The asymptotic flatness at spatial infinity requires that 
\be
f(r \to \infty)=1\,,\qquad h(r \to \infty)=1\,,\qquad
\phi'(r \to \infty)=0\,,\qquad V( \phi_{\infty})=0\,,
\label{bouninf}
\ee
where $\phi_{\infty} \equiv \phi (r \to \infty)$. 
For the power-law potential $V(\phi)=\lambda_n \phi^n$,
the field value $\phi_{\infty}$ is equivalent to 0. 
In this case, the nonminimal coupling (\ref{Fphi}) approaches 
the value 1 of GR in the limit $r \to \infty$. 
For the massless scalar field without the potential, we impose the 
boundary condition $\phi_{\infty}=0$ besides 
the first three of (\ref{bouninf}).
Since only the ratio between $f'$ and $f$ appears in Eqs.~(\ref{macon}) 
and (\ref{eq1})-(\ref{eq3}), 
the constant $f_0$ in the expansion of Eq.~(\ref{fexpan}) 
can be chosen as any arbitrary constant.
The asymptotic value of $f$ at spatial infinity is generally different 
from 1, but it can be shifted to 1 by the time reparametrization. 
For $M={\rm constant}$, Eq.~(\ref{hr}) shows that the function $h$ approaches 1 
as $r \to \infty$.
The field value $\phi_0$ at $r=0$ can be determined by a shooting method 
to satisfy the boundary conditions (\ref{bouninf}) at spatial infinity.

For the numerical purpose, we introduce the density $\tilde{\rho}_0$ 
and the distance $r_0$, as
\ba
\tilde{\rho}_0
&=& m_n n_0=1.6749 \times 10^{14}~{\rm g} \cdot {\rm cm}^{-3}\,,\\
r_0 
&=& \frac{c}{\sqrt{G \tilde{\rho}_0}}=89.664~{\rm km}\,,
\ea
where $m_n=1.6749 \times 10^{-24}$~g is the neutron mass 
and $n_0=0.1~{\rm (fm)}^{-3}$ is the typical number 
density of NSs. 
It is convenient to define the following dimensionless variables
\be
y \equiv \frac{\rho}{\tilde{\rho}_0}\,,\qquad 
z \equiv \frac{P}{\tilde{\rho}_0}\,,\qquad 
v \equiv \frac{V}{\tilde{\rho}_0}\,,\qquad
v_{,\varphi} \equiv  \frac{M_{\rm pl}V_{,\phi}}{\tilde{\rho}_0}\,,\qquad
m \equiv \frac{3{\cal M}}{4\pi r_0^3 \tilde{\rho}_0}\,,\qquad 
\varphi \equiv \frac{\phi}{M_{\rm pl}}\,,\qquad 
s \equiv \ln \frac{r}{r_0}\,.
\label{dimen}
\ee
Then, the quantities $\xi$ and $\zeta$ in Eq.~(\ref{xizeta})
are expressed, respectively, as 
\be
\xi=\alpha_1+\alpha_2 \ln y\,,\qquad 
\zeta=\alpha_3+\alpha_2 \ln z\,,
\ee
where $\alpha_1=\ln(\tilde{\rho}_0/{\rm g\cdot cm^{-3}})/\ln10$, 
$\alpha_2=(\ln10)^{-1}$, 
and $\alpha_3=\ln(\tilde{\rho}_0\,c^2/{\rm dyn\cdot cm^{-2}})/\ln10$. 
Then, the EOS translates to the form 
\be
z=\exp \left[ \frac{\zeta(\xi)-\alpha_3}{\alpha_2} \right]\,,
\label{zy}
\ee
where $\zeta(\xi)$ is the function on the right hand side 
of Eq.~(\ref{zeta}). 
{}From the continuity Eq.~(\ref{macon}), 
the derivative $y_{,s} \equiv \rd y/\rd s$ is expressed as 
\be
y_{,s}
=-\frac{y(y+z)}{2z} \left( \frac{{\rm d} \zeta}
{{\rm d} \xi} \right)^{-1}
\frac{f_{,s}}{f}\,.
\label{ys}
\ee
{}From Eqs.~(\ref{eq1})-(\ref{eq3}), we have
\ba
\hspace{-0.8cm}
& &
\frac{f_{,s}}{f}
=-\frac{2(h-1)-16\pi e^{2s+2Q\varphi} (z-v)
+h[(6Q^2-1) \varphi_{,s}-8Q] \varphi_{,s}}
{2h (1-Q \varphi_{,s})}\,,\label{fs}\\
\hspace{-0.8cm}
& &
m_{,s}
= \left[ 16 \pi  (1-Q \varphi_{,s}) \right]^{-1} 
[ 3e^s (1+2Q^2) \varphi_{,s}^2-8\pi  m\,\varphi_{,s} 
\{\varphi_{,s}+2Q(Q \varphi_{,s}-1) \} \nonumber \\
\hspace{-0.8cm}
& &\qquad +48\pi e^{3s+2Q\varphi} \{ (8Q^3\varphi_{,s}-8Q^2
-2Q\varphi_{,s}+1)v+(Q\varphi_{,s}-1)(2Q^2y -y+2Qv_{,\varphi})
+Q(6Q-6Q^2\varphi_{,s}+\varphi_{,s})z \} ],\\
\hspace{-0.8cm}
& &
\varphi_{,ss}=-[1 +4\pi e^{2s+2Q\varphi} \{ 2Q^2 
(y-3z+4v)-2v+2Qv_{,\varphi}-y+z \}] \frac{\varphi_{,s}}{h}+
\frac{8\pi}{h}[Q(y-3z+4v)+v_{,\varphi}]e^{2s+2Q\varphi}\,,
\label{varphis}
\ea
with 
\be
h=1-\frac{8\pi m}{3e^s}\,.
\label{hx}
\ee
Solving Eqs.~(\ref{zy})-(\ref{hx}) with the boundary conditions 
(\ref{fexpan})-(\ref{Pexpan}) outwards, we know the values of $y$, $z$, $f$, 
$m$, and $\varphi$ inside the star. 
Outside the star, we can simply set $y=0=z$ and solve 
Eqs.~(\ref{fs})-(\ref{hx}) for  $f$, $m$, and $\varphi$. 
Defining $m_{\infty} \equiv m(r \to \infty)$, the ADM mass $M$ 
of star can be computed as 
\be
M=2.5435 \times 10^2\,m_{\infty}\,M_{\odot}\,,
\ee
where $M_{\odot}=1.9884 \times 10^{33}$~g is the solar mass.

\subsection{Einstein frame}

Under the so-called conformal transformation 
\be
(g_{\mu \nu})_{\rE}=F(\phi) g_{\mu \nu}\,,
\ee
the action (\ref{actionJ}) can be transformed to that 
in the Einstein frame without the nonminimal coupling \cite{Fujii}. 
Here and in the following, we use the roman subscript  ``$\rE$'' 
to represent quantities in the Einstein frame.
The Ricci scalars in two frames are related to 
each other, as $R=(R_{\rE}-6g^{\mu \nu}_{\rE} \partial_{\mu} 
\omega \partial_{\nu} \omega+6 \square_{\rE}\,\omega)F$, 
where $\omega=(1/2) \ln F=-Q\phi/M_{\rm pl}$ and 
$\square_{\rE}\omega=(1/\sqrt{-g_{\rE}})
\partial_{\mu} (\sqrt{-g_{\rE}}\,g^{\mu \nu}_{\rE}
\partial_{\nu} \omega)$. 
On using the property $\sqrt{-g}=F^{-2}\sqrt{-g_{\rE}}$ 
and dropping a boundary term associated with 
$\square_{\rE}\omega$, the action (\ref{actionJ}) reduces to 
\be
{\cal S}_{\rE}=\int \rd^4 x \sqrt{-g_{\rE}} \left[ 
\frac{M_{\rm pl}^2}{2} R_{\rE} -\frac{1}{2} 
g^{\mu \nu}_{\rE} \partial_{\mu} \phi 
\partial_{\nu} \phi-V_{\rE}(\phi) \right]
+\int \rd^4 x\,{\cal L}_m \left( F^{-1}(\phi) 
(g_{\mu \nu})_{\rE}, \Psi_m \right)\,,
\label{actionE}
\ee
where
\be
V_{\rE} (\phi)=\frac{V(\phi)}{F^2(\phi)}\,.
\ee

From Eq.~(\ref{actionE}), it is clear that the canonical scalar field 
$\phi$ is directly coupled to matter fields in the Einstein frame.
The matter energy-momentum tensor in the Einstein frame, 
which is defined by 
$(T_{\mu \nu})_{\rE}=-(2/\sqrt{-g_{\rE}}) \delta {\cal L}_m/ 
\delta g^{\mu \nu}_{\rE}$, is related to that in the Jordan frame, 
as $(T_{\mu \nu})_{\rE}=T_{\mu \nu}/F$. 
The energy density $\rho_{\rE}$ and the pressure $P_{\rE}$ 
of perfect fluids in the Einstein frame is given by 
$(T^{\mu}_{\nu})_{\rE}={\rm diag} (-\rho_{\rE}, 
P_{\rE}, P_{\rE}, P_{\rE})$, so there are the following 
relations 
\be
\rho_{\rE}=\frac{\rho}{F^2}\,,\qquad 
P_{\rE}=\frac{P}{F^2}\,.
\label{rhoE}
\ee
Varying the action (\ref{actionE}) with respect to $\phi$ and 
using the relation 
$\partial {\cal L}_m/\partial \phi=-\sqrt{-g_{\rE}}\,T_{\rE}\,F_{,\phi}/(2F)$, 
where $T_{\rE}=-\rho_{\rE}+3P_{\rE}$ is the trace of energy-momentum tensor, 
it follows that 
\be
\square_{\rE} \phi-V_{{\rE},\phi}-\frac{Q}{M_{\rm pl}} 
\left( \rho_{\rE}-3P_{\rE} \right)=0\,.
\label{phiEin}
\ee
This shows that the matter coupling $Q$ modifies the dynamics 
of $\phi$.

In the Einstein frame, we consider the spherically symmetric and static 
background given by the line element  
\be
{\rm d}s_{\rE}^2=-f_{\rE}(r_{\rE}) \rd t^2+h_{\rE}^{-1}(r_{\rE}) \rd r_{\rE}^2 
+r_{\rE}^2 \left( \rd \theta^2+\sin^2 \theta\,\rd \varphi^2 \right)\,.
\label{metE}
\ee
Since ${\rm d}s_{\rE}^2=F\rd s^2$, the distance $r$ and 
the metrics $f, h$ in the Jordan frame 
are related to those in the Einstein frame, as
\ba
r &=&  e^{Q\phi/M_{\rm pl}} r_{\rE}\,,\\
f(r) &=& e^{2Q\phi/M_{\rm pl}} f_{\rE} (r_{\rE})\,,\\
h(r) &=& h_{\rE} (r_{\rE}) \left( 1+\frac{Qr_{\rE}}{M_{\rm pl}}
\frac{\rd \phi}{\rd r_{\rE}} \right)^2\,.
\label{hcore}
\ea
We introduce the mass function ${\cal M}_{\rE}  (r_{\rE})$ 
in the Einstein frame, as
\be
h_{\rE} (r_{\rE})=1-\frac{2G{\cal M}_{\rE} (r_{\rE})}{r_{\rE}}\,,
\label{hrE}
\ee
together with the asymptotic mass 
\be
M_{\rE} \equiv \lim_{r_{\rE} \to \infty} {\cal M}_{\rE}(r_\rE)=
\frac{r_{\rE}}{2G} \left( 1-h_{\rE} \right) \biggl|_{r_{\rE} \to \infty}\,.
\ee
In the Einstein frame, the star radius $r_s$ corresponds to 
\be
(r_{s})_{\rE}=e^{-Q \phi_s/M_{\rm pl}}r_s\,,
\ee
where $\phi_s$ is the field value at the surface of star.

On using the correspondence (\ref{hcore}) with Eqs.~(\ref{hr}) and 
(\ref{hrE}), it follows that 
\be
{\cal M} (r)=e^{Q\phi/M_{\rm pl}} \left[ {\cal M}_{\rE} (r_{\rE})
-4\pi M_{\rm pl}Q  r_{\rE}^2 \frac{\rd \phi}{\rd r_{\rE}} 
\left( 2+\frac{Qr_{\rE}}{M_{\rm pl}}
\frac{\rd \phi}{\rd r_{\rE}} \right) 
\left( 1-\frac{2G{\cal M}_{\rE}(r_{\rE})}{r_{\rE}} \right) \right]\,.
\label{Mtra}
\ee
The existence of terms $e^{Q\phi/M_{\rm pl}}$ and $\rd \phi/\rd r_{\rE}$ 
lead to the difference between ${\cal M}(r)$ and ${\cal M}_{\rE} (r_{\rE})$. 
If $\rd \phi/\rd r_{\rE}$ decreases faster than $1/r_{\rE}^2$ 
and $\phi$ approaches 0 as $r_{\rE} \to \infty$, 
then we have $M=M_{\rE}$.
On the other hand, if the field at large distances
has the radial dependence 
\be
\frac{\rd \phi}{\rd r_{\rE}}=\frac{\alpha}{r_{\rE}^2}\,,
\label{dphi}
\ee
where $\alpha$ is a constant, it follows that 
\be
M=e^{Q \phi_{\infty}/M_{\rm pl}} 
 \left( M_{\rE}-8\pi M_{\rm pl} Q \alpha 
 \right)\,.
\label{MME}
\ee
Even when $\phi_{\infty}=0$, the nonvanishing radial derivative (\ref{dphi}) leads to 
the difference between $M$ and $M_{\rE}$. 
As we will discuss in Sec.~\ref{masslesssec}, this difference appears 
for BD theories with $V(\phi)=0$.

In the Einstein frame, the matter continuity Eq.~(\ref{macon}) reads 
\be
\frac{\rd P_{\rE}}{\rd r_{\rE}}+\frac{1}{2f_{\rE}}
\frac{\rd f_{\rE}}{\rd r_{\rE}} \left( \rho_{\rE}+P_{\rE} 
\right)+\frac{Q}{M_{\rm pl}} \left( \rho_{\rE}-3P_{\rE} 
\right) \frac{\rd \phi}{\rd r_{\rE}}=0\,, 
\label{PEeq}
\ee
while the scalar-field Eq.~(\ref{phiEin}) reduces to 
\be
\frac{\rd^2 \phi}{\rd r_{\rE}^2}+\left[ \frac{2}{r_{\rE}}
+\frac{1}{2} \frac{\rd}{\rd r_{\rE}} \ln \left( f_{\rE} h_{\rE} 
\right) \right] 
\frac{\rd \phi}{\rd r_{\rE}}-\frac{1}{h_{\rE}} \left[ 
V_{\rE, \phi}+\frac{Q}{M_{\rm pl}} 
\left( \rho_{\rE}-3P_{\rE} \right) \right]=0\,.
\label{phiEeq}
\ee
Varying the action (\ref{actionE}) with respect to $(g_{\mu \nu})_{\rE}$, 
the metric $f_{\rE}$ and the mass function ${\cal M}_{\rE}$ obey 
\ba
\frac{1}{f_{\rE}}\frac{\rd f_{\rE}}{\rd r_{\rE}}
&=& -\frac{1}{2h_{\rE}r_{\rE}M_{\rm pl}^2}
\left[ 2M_{\rm pl}^2 \left( h_{\rE}-1 \right)
-r_{\rE}^2 \left\{ 2P_{\rE}-2V_{\rE}+h_{\rE} 
\left( \frac{\rd \phi}{\rd r_{\rE}} \right)^2 \right\} 
\right]\,, \\
\frac{\rd {\cal M}_{\rE}}{\rd r_{\rE}}
&=& 4\pi r_{\rE}^2 \left[ \rho_{\rE}+V_{\rE}
+\frac{h_{\rE}}{2} 
\left( \frac{\rd \phi}{\rd r_{\rE}} \right)^2 
\right]\,,
\label{MEeq}
\ea
which show that the field potential $V_{\rE}$ and  
the kinetic energy $(\rd \phi/\rd r_{\rE})^2$ modify the values 
of $f_{\rE}$ and ${\cal M}_{\rE}$ in GR. 
{}From Eqs.~(\ref{PEeq}) and (\ref{phiEeq}), we find that  
the field $\phi$ and matter interact with each other through the coupling $Q$.
While the equations of motion in the Einstein frame are simpler than those 
in the Jordan frame, the EOS (\ref{zeta}) needs to be transformed 
to the relation between $P_{\rE}$ and $\rho_{\rE}$. 
The boundary conditions at $r_{\rE}=0$ and $r_{\rE} \to \infty$
are similar to those in the Jordan frame, i.e., 
\be
\frac{\rd f_{\rE}}{\rd r_{\rE}} (r_{\rE}=0)=0\,,\qquad 
\frac{\rd h_{\rE}}{\rd r_{\rE}} (r_{\rE}=0)=0\,,
\qquad \frac{\rd \phi}{\rd r_{\rE}} (r_{\rE}=0)=0\,,\qquad 
\frac{\rd \rho_{\rE}}{\rd r_{\rE}} (r_{\rE}=0)=0\,,
\label{bounE}
\ee
and 
\be
f_{\rE}(r_{\rE} \to \infty)=1\,,\qquad h_{\rE}(r_{\rE} \to \infty)=1\,,\qquad
 \frac{\rd \phi}{\rd r_{\rE}} (r_{\rE} \to \infty)=0\,,
\qquad V_{\rE}( \phi_{\infty})=0\,.
\label{bouninfE}
\ee
The analytic solutions to $f_{\rE}, h_{\rE}, \phi, P_{\rE}$ expanded around 
$r_{\rE}=0$ can be obtained in a similar way to those derived 
in Eqs.~(\ref{fexpan})-(\ref{Pexpan}) in the Jordan frame. 
The resulting solutions consistent with Eqs.~(\ref{PEeq})-(\ref{MEeq}) are given by 
\ba
f_{\rm E}(\re)&=&f_{\rE 0} 
\left[1+\frac{\rho_{\rE 0} +3P_{\rE 0} -2V_{\rE}(\phi_{0}) }{6\Mpl^2}\re^2
+{\cal O}(\re^4)\right]\,,\label{fexpanE}\\
h_{\rm E}(\re)&=&
1-\frac{\rho_{\rE 0}+V_{\rE}(\phi_0)}{3\Mpl^2}\re^2
+{\cal O}(\re^4)\,,\label{hexpanE}\\
\phi(\re)&=&
\phi_0+\frac{\Mpl V_{\rE,\phi}(\phi_0)+Q(\rho_{\rE0}-3P_{\rE0})}{6\Mpl}\re^2
+{\cal O}(\re^4)\,, 
\label{phiEinb}\\
P_{\rm E}(\re)&=&P_{\rE0}
-\frac{(\rho_{\rE0}+P_{\rE0})\{\rho_{\rE0}+3P_{\rE0}-2V_{\rE}(\phi_0)\}
+2Q(\rho_{\rE0}-3P_{\rE0})\{Q(\rho_{\rE0}-3P_{\rE0})+\Mpl V_{\rE,\phi}(\phi_0)\}}{12\Mpl^2}\re^2
\notag\\
&&+{\cal O}(\re^4)\,,\label{PexpanE} 
\ea
where $f_{\rE0},\rho_{\rE0},P_{\rE0}$ are constants corresponding to 
$f_0,\rho_0,P_0$ in Eq.~(\ref{exp}), respectively. 
The field value $\phi_0$ is iteratively known to satisfy the boundary 
conditions  (\ref{bouninfE}) at spatial infinity.
Solving Eqs.~(\ref{PEeq})-(\ref{MEeq}) numerically 
and transforming the solutions back to the Jordan frame, 
the physical observables like $r_s$ and $M$ should coincide
with those computed directly in the Jordan frame for given model parameters.
We will address this issue in Sec.~\ref{masslesssec}.

%%%%%%%%%%%%%%%%%%%%%%%%%%%%%%
\section{Massless Brans-Dicke theories}
\label{masslesssec}
%%%%%%%%%%%%%%%%%%%%%%%%%%%%%%

We first consider massless BD theories without the 
scalar potential, i.e., 
\be
V(\phi)=0\,.
\ee
In this case, we can set $v=0$ and $v_{,\varphi}=0$ in 
Eqs.~(\ref{fs})-(\ref{varphis}).
We recall that the EOS is written as the form (\ref{ys}), 
with $\zeta(\xi)$ and $z$ given by 
Eqs.~(\ref{zeta}) and (\ref{zy}), respectively.
Numerically, we solve Eqs.~(\ref{fs})-(\ref{varphis}) with 
Eqs.~(\ref{zy}), (\ref{ys}), and (\ref{hx}) 
by using the boundary conditions (\ref{fexpan})-(\ref{Pexpan}) around $r=0$.

{}From Eqs.~(\ref{phiexpan}) and (\ref{Pexpan}), the scalar field 
and pressure around $r=0$ are given, respectively, by 
\ba
\phi(r)
&=& \phi_0+\frac{Q(\rho_0-3P_0)}{6M_{\rm pl}
e^{-2Q \phi_0/M_{\rm pl}}}r^2+{\cal O} (r^4)\,,
\label{phiana}\\
P(r)
&=& P_0-\frac{(1+2Q^2)\rho_0+3(1-2Q^2)P_0} 
{12M_{\rm pl}^2 e^{-2Q \phi_0/M_{\rm pl}}}
\left( \rho_0+P_0 \right)r^2+{\cal O} (r^4)\,.
\label{Pexpansion}
\ea
%

%%%%%%%%%%%%%%%%%%%%%%%%%%%%%%
\begin{figure}[h]
\begin{center}
\includegraphics[height=3.3in,width=3.5in]{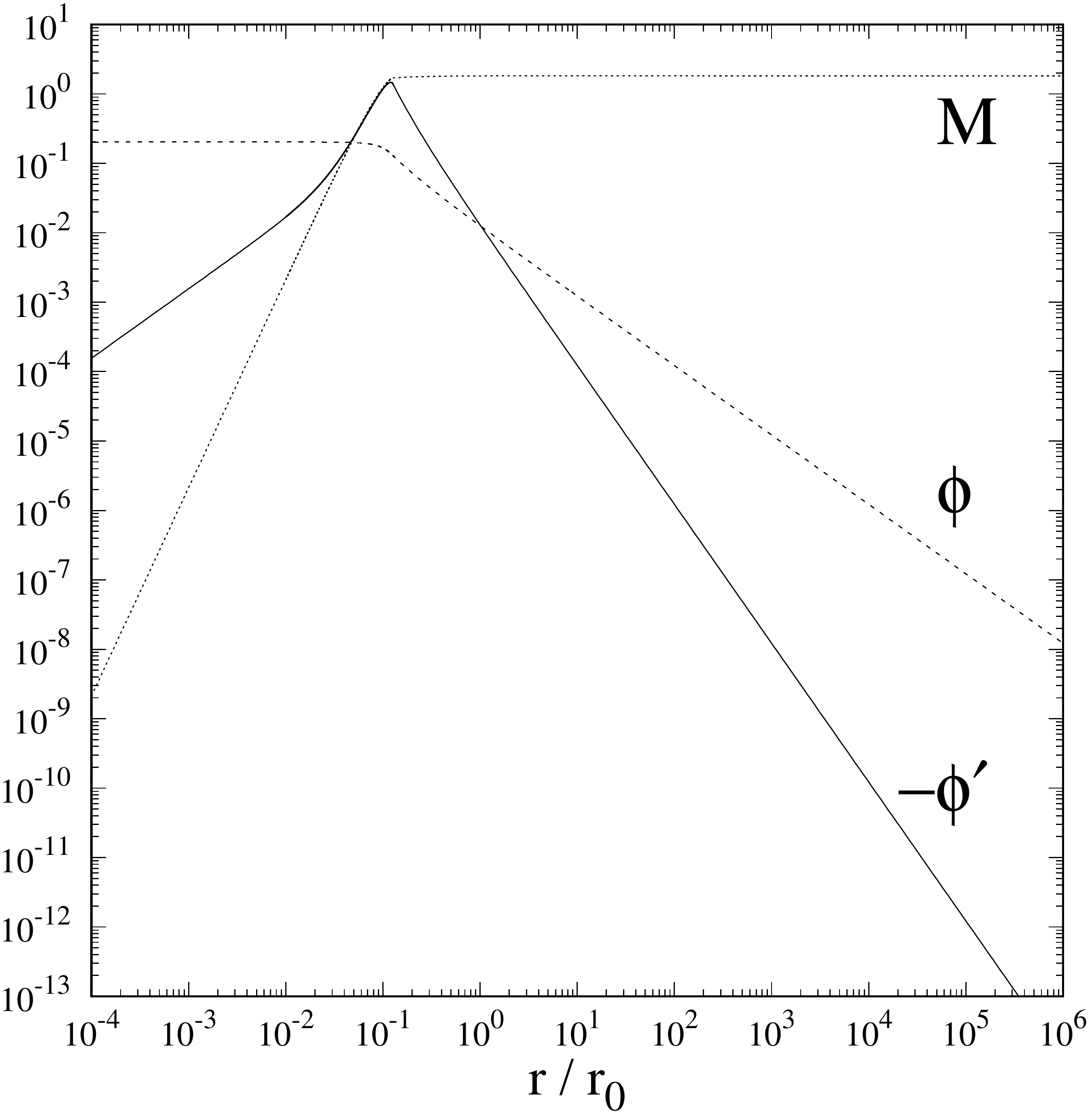}
\includegraphics[height=3.3in,width=3.5in]{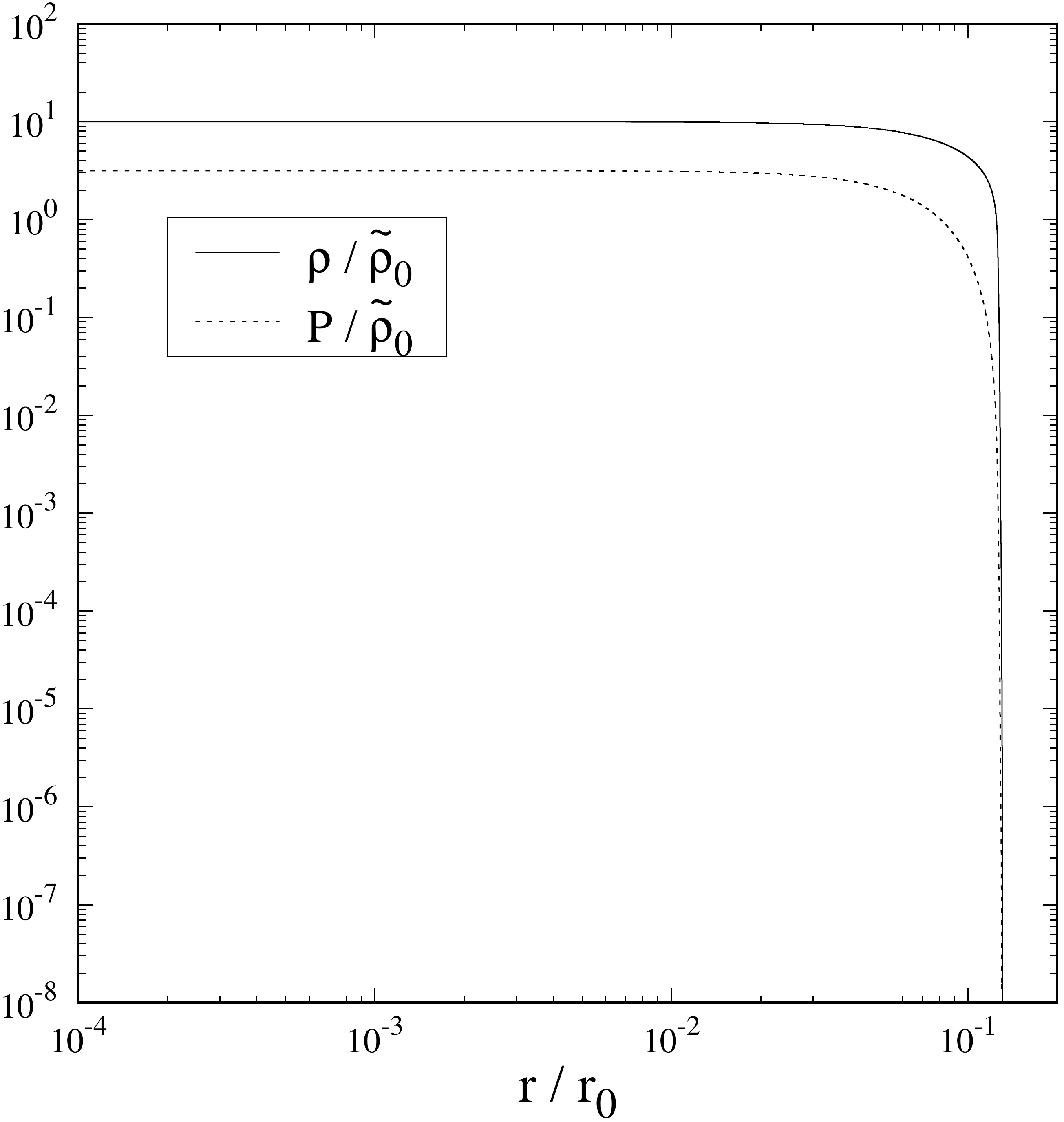}
\end{center}
\caption{\label{fig1}
(Left) $\phi$, $-\phi'$, and ${\cal M}$ (normalized by 
$M_{\rm pl}$, $r_0 M_{\rm pl}$, and $M_{\odot}$, respectively) 
versus $r/r_0$ inside and outside the NS 
for BD theories with $Q=-1/\sqrt{6}$ and $V(\phi)=0$. 
We adopt the SLy EOS with the central density 
$\rho_0=10\tilde{\rho}_0=1.6749 \times 10^{15}$~g\,$\cdot$\,cm$^{-3}$. 
We choose the boundary conditions 
(\ref{fexpan})-(\ref{Pexpan}) 
at $s=\ln (r/r_0)=-10$, with $\phi_0$ giving rise to the 
asymptotic values $\phi (r \to \infty)=0$ and 
$\phi'(r \to \infty)=0$.
(Right) $\rho/\tilde{\rho}_0$ and 
$P/\tilde{\rho}_0$ versus the distance $r/r_0$ inside the NS 
for the same model parameters and boundary conditions as those used in the left panel.}
\end{figure}
%%%%%%%%%%%%%%%%%%%%%%%%%%%%%%

If $Q=0$, then the scalar field stays constant ($\phi=\phi_0$) 
around the center of star.
Indeed, this can be also confirmed by the field Eq.~(\ref{phiEeq})
in the Einstein frame. 
For $Q=0$, the general solution to Eq.~(\ref{phiEeq}) 
is expressed in the integrated form 
\be
\phi(r_{\rE})=\phi_0+\alpha \int_{0}^{r_{\rE}} \frac{1}{\tilde{r}_{\rE}^2 
\sqrt{f_{\rE}h_{\rE}}} \rd \tilde{r}_{\rE}\,,
\label{phirE}
\ee
where $\alpha$ is a constant.
To avoid the divergence of the integral in Eq.~(\ref{phirE}) 
around $r_{\rE}=0$, we require that $\alpha=0$ and hence 
$\phi$ is constant at any radial distance $r_{\rE}$.

For $Q \neq 0$, the scalar field varies with the increase of $r$. 
In the following, we consider the negative coupling 
\be
Q<0\,,
\ee
without loss of generality.
We also focus on the coupling in the range $Q^2 \leq 1/2$, 
under which the pressure (\ref{Pexpansion}) decreases 
with the growth of $r$.
Our analysis covers dilaton gravity ($Q=-1/\sqrt{2}$) 
as a special case. 
The massless BD theory with $Q=-1/\sqrt{6}$, which we 
discuss in this section, is different from 
$f(R)$ gravity, in that the latter contains the 
nonvanishing potential $V(\phi)$. 
We will study the BD theory with $V(\phi) \neq 0$ in 
Secs.~\ref{massivesec} and \ref{selfsec}.

%%%%%%%%%%%%%%%%%%%%%%%%%%%%%%
\begin{figure}[h]
\begin{center}
\includegraphics[height=3.4in,width=3.5in]{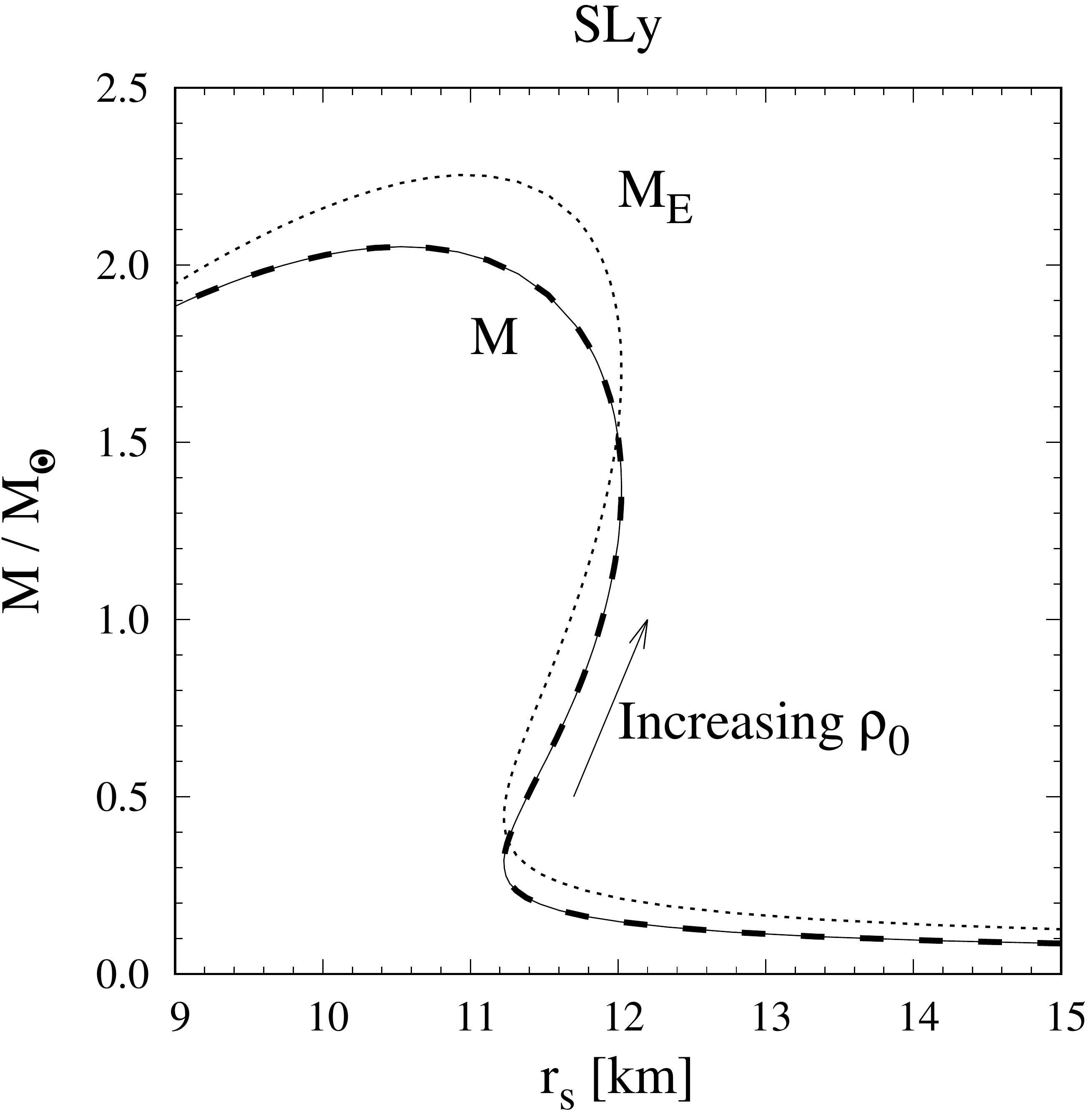}
\includegraphics[height=3.4in,width=3.5in]{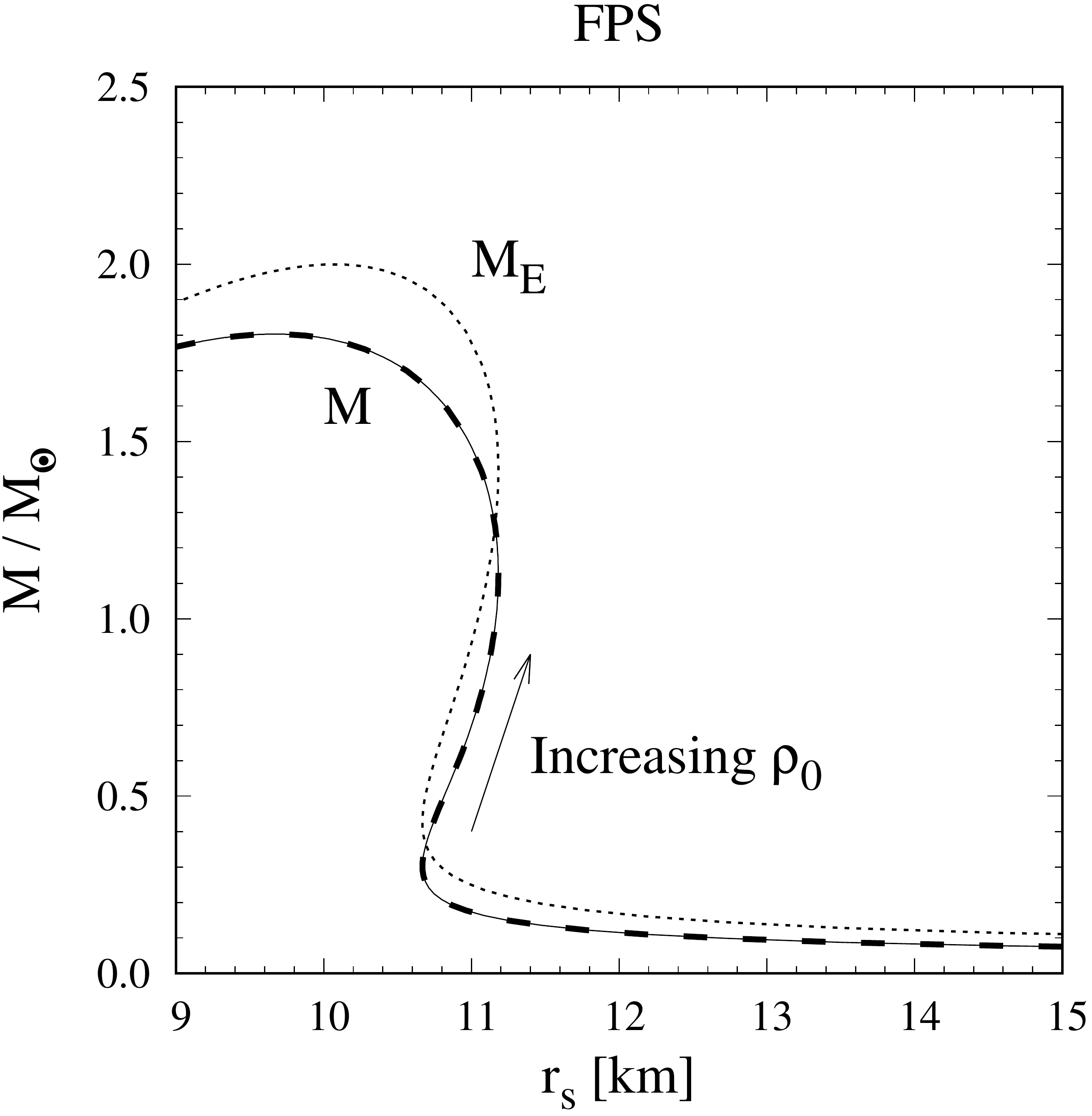}
\end{center}
\caption{\label{fig2}
Mass-radius relation for the SLy (left) and FPS (right) EOSs 
in BD theories with $Q=-1/\sqrt{6}$ and $V(\phi)=0$. 
We show the masses $M$ and $M_{\rm E}$ (both are 
normalized by $M_{\odot}$) computed in 
the Jordan and Einstein frames, respectively. 
The bold dashed lines correspond to the mass $M$ obtained from 
$M_{\rm E}$ by using the transformation from the Einstein 
frame to the Jordan frame.
}
\end{figure}
%%%%%%%%%%%%%%%%%%%%%%%%%%%%%%

%%%%%%%%%%%%%%%%%%%%%%%%%%%%%%
\begin{figure}[h]
\begin{center}
\includegraphics[height=3.4in,width=3.5in]{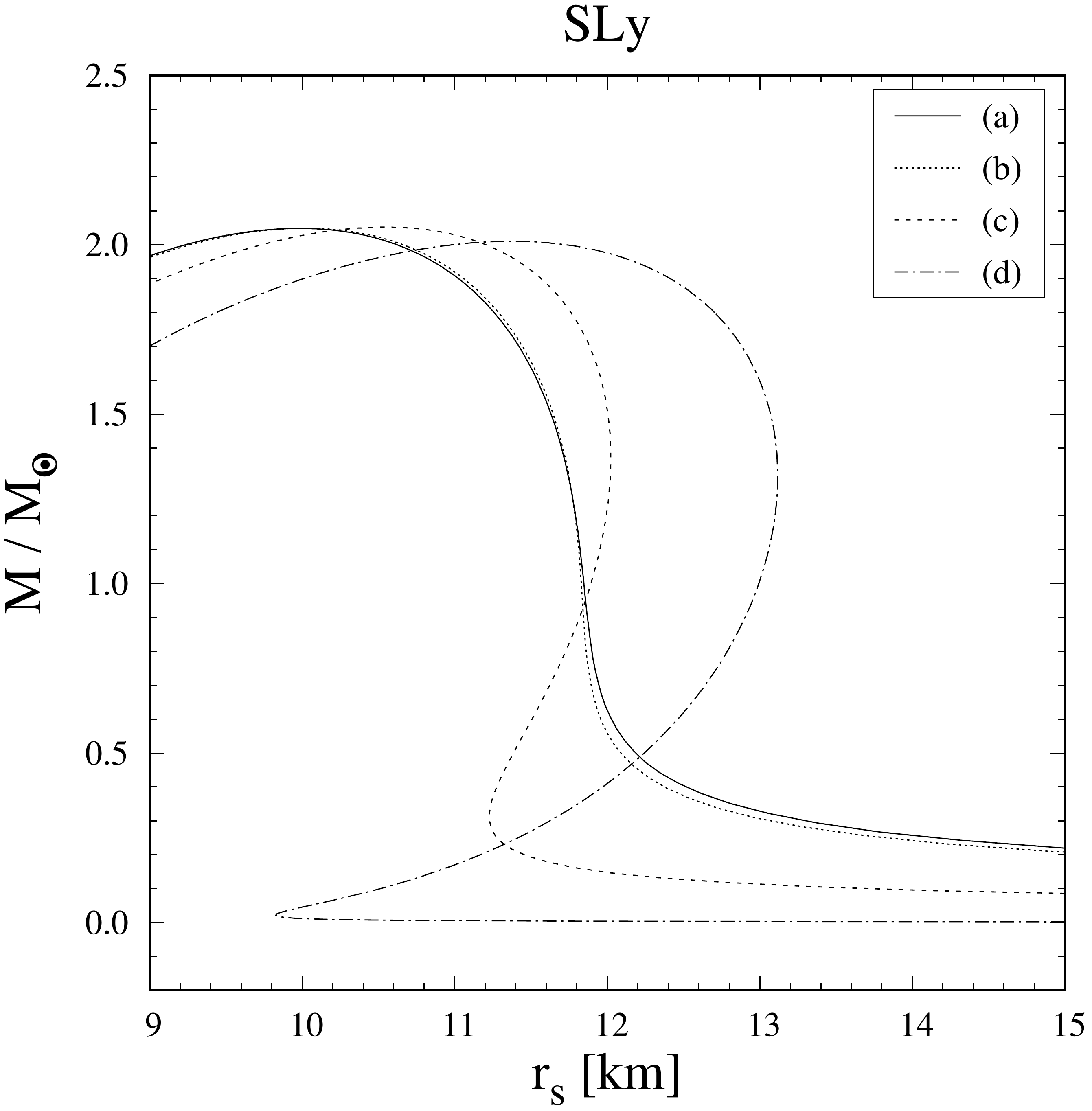}
\includegraphics[height=3.4in,width=3.5in]{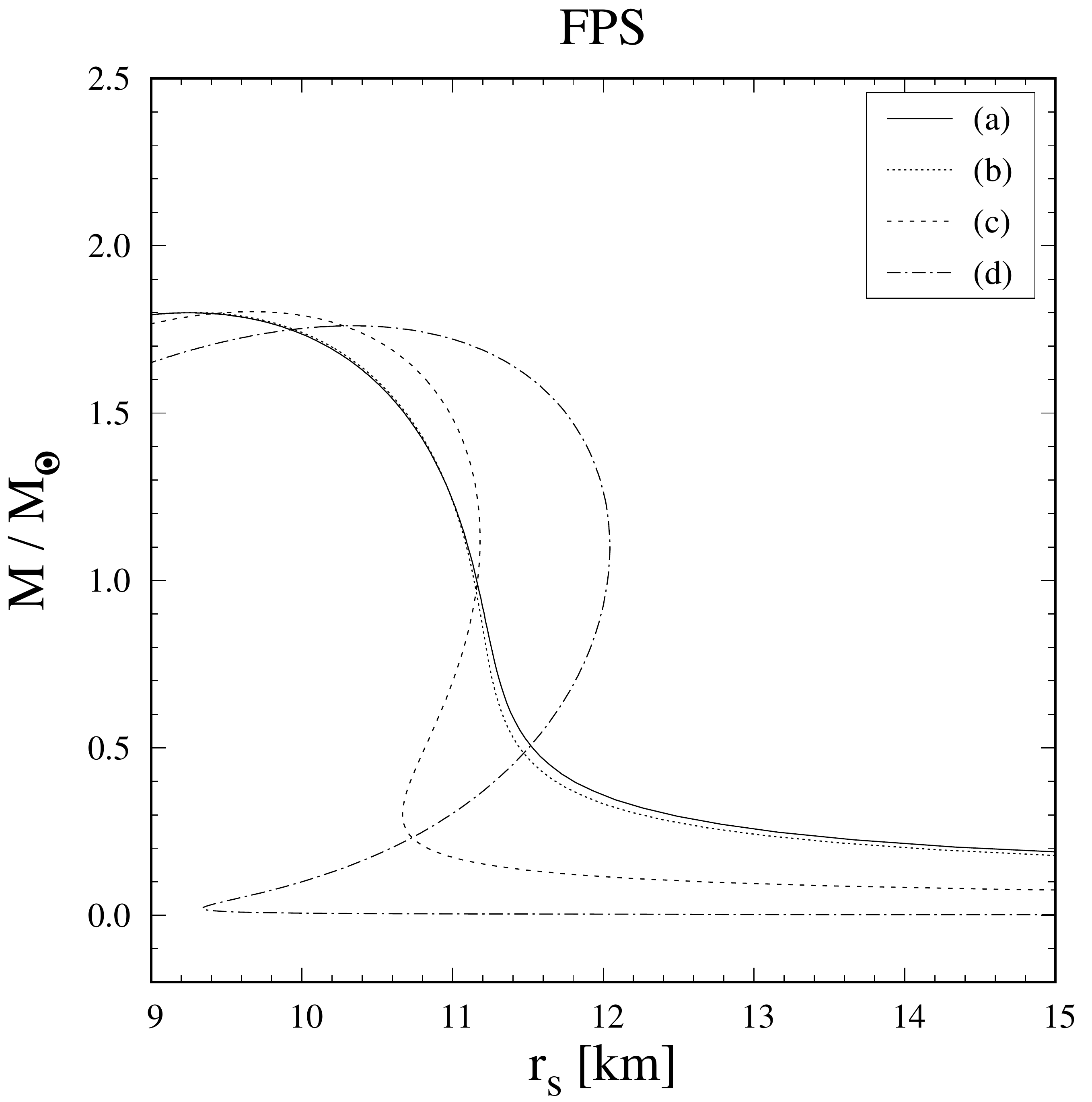}
\end{center}
\caption{\label{fig3}
Mass-radius relation computed in the Jordan frame 
for the SLy (left) and FPS (right) EOSs 
in BD theories with $V(\phi)=0$.
Each line corresponds to
(a) $Q=0$ (GR), (b) $Q=-0.1$, (c) $Q=-1/\sqrt{6}$, 
and (d) $Q=-1/\sqrt{2}$.
}
\end{figure}
%%%%%%%%%%%%%%%%%%%%%%%%%%%%%%

Under the condition $\rho_0>3P_0$, 
Eq.~(\ref{phiana}) shows that $\phi'(r)<0$ around $r=0$. 
In the left panel of Fig.~\ref{fig1}, we plot $\phi(r)$, $-\phi'(r)$, 
and ${\cal M}(r)$ versus $r/r_0$ for the SLy EOS 
with the central density $\rho_0$ satisfying the condition 
$Q(\rho_0-3P_0)<0$. 
In this case, $-\phi'(r)$ is positive around $r=0$ and it 
linearly grows as $-\phi'(r) \propto r$ for $r \lesssim 0.01r_0$. 
In this regime, the field $\phi$ slowly decreases from the central 
value $\phi_0 \simeq 0.203M_{\rm pl}$ according to Eq.~(\ref{phiana}). 
The field derivative $-\phi'(r)$ reaches the maximum around the surface of 
star ($r_s \simeq 0.13r_0$). In the left panel of Fig.~\ref{fig1}, 
we observe that the large variation of $\phi$ starts to 
occur around $r=r_s$. 
In the right panel, we also find that both $\rho$ and $P$ rapidly 
drop down for $r>0.1 r_0$.

Outside the star, both $\phi (r)$ and $-\phi'(r)$ 
decrease with the increase of $r$.
Since $\rho_{\rE}=0=P_{\rE}$ for $r>r_s$, the solution to 
Eq.~(\ref{phiEeq}) is the same as Eq.~(\ref{phirE}). 
For the large distance $r \gg r_s$, both $f_{\rE}$ and 
$h_{\rE}$ approach 1 with $\phi_{\infty} \to 0$ and 
$r_{\rE} \to r$, so the field derivative is given by 
\be
\phi' (r) \simeq \frac{\alpha}{r^2}\,.
\label{phid}
\ee
In Fig.~\ref{fig1}, we can confirm that $-\phi'(r)$ decreases in 
proportion to $1/r^2$ for $r \gg r_s$, with $\phi(r)$ 
approaching 0 at spatial infinity.
For the massless scalar, the field value $\phi_0$ at $r=0$ 
satisfying the boundary condition $\phi_{\infty}=0$ 
can be identified in the following way. 
First, we perform the numerical integration by choosing 
$\phi_0=0$ and then find the asymptotic value $\phi_{\rm asy}$ 
at $r \gg r_s$ (say, at $r=10^{20}r_s$). 
Then, we run the code again with the value $\phi_0=-\phi_{\rm asy}$ 
at $r=0$. This second run leads to the asymptotic value  
$\phi_{\infty}$ converging to 0. 
It is important to identify the appropriate value of $\phi_0$ 
in this way because the mass $M$ and radius 
$r_s$ are affected by the nonminimal coupling term 
$e^{-2Q \phi_0/M_{\rm pl}}$. 
In the numerical simulation of Fig.~\ref{fig1}, the mass function 
${\cal M}$ quickly approaches the asymptotic 
value $M=1.83 M_{\odot}$ for
$r>r_s=0.13r_0=11.7$ km.

In Fig.~\ref{fig2}, we show the mass $M$ in the 
Jordan frame versus the radius $r_s$ for the SLy (left) 
and FPS (right) EOSs with 
the coupling $Q=-1/\sqrt{6}$. As 
the central density $\rho_0$ grows from the value of 
order $10^{14}$~g\,$\cdot$\,cm$^{-3}$, the mass $M$ increases by 
reaching a maximum $M_{\rm max}$. 
For SLy, we have $M_{\rm max}\simeq 2.05 M_{\odot}$ 
with the radius $r_s \simeq 10.5$~km around 
the density $\rho_0=2.7 \times 10^{15}$~g\,$\cdot$\,cm$^{-3}$.
For this maximum mass the condition $\rho_0>3P_0$ holds, 
but as $\rho_0$ increases further, the system eventually 
enters the region with the fully relativistic EOS 
satisfying $\rho_0<3P_0$. 
With the growth of $\rho_0$ in the fully relativistic region, 
the radius $r_s$ gets smaller,  
by reflecting the fact that the second term on the right hand side of Eq.~(\ref{Pexpansion}) increases. 
This is also accompanied by the decrease of $M$.
For $Q=-1/\sqrt{6}$, the maximum mass exceeds $2M_{\odot}$ 
with the SLy EOS, while, the FPS EOS gives rise to the 
maximum mass $M_{\rm max}\simeq 1.80 M_{\odot}$ with 
$r_s \simeq $~9.67 km and 
$\rho_0=3.5 \times 10^{15}$~g\,$\cdot$\,cm$^{-3}$.

In Fig.~\ref{fig2}, we also plot the mass $M_{\rE}$ computed in the Einstein frame. 
As we estimated in Eq.~(\ref{MME}), there is the 
relation $M=M_{\rE}-8\pi M_{\rm pl}Q \alpha$ for 
$\phi_{\infty}=0$, where 
$\alpha$ corresponds to the coefficient in Eq.~(\ref{phid}).
The constant $\alpha$ is related to the field derivative 
$\phi'(r)$ at $r=r_s$. If we extrapolate the solution (\ref{phiana}) up to the radius of star, 
it follows that $\phi'(r_s) \approx Q(\rho_0-3P_0)r_s
/(3M_{\rm pl} e^{-2Q\phi_0/M_{\rm pl}})$. 
If the exterior solution (\ref{phid}) at spatial infinity 
is also extrapolated down to $r=r_s$, then 
the coefficient $\alpha$ can be estimated as 
$\alpha \approx Q(\rho_0-3P_0)r_s^3
/(3M_{\rm pl} e^{-2Q\phi_0/M_{\rm pl}})$. 
Although this is a crude estimation under which the 
coefficient $\alpha$ is inaccurate, we may 
generally express $\alpha$ in the form 
\be
\alpha=\beta \frac{Q\rho_0 r_s^3}{M_{\rm pl}}\,,
\label{aldef}
\ee
where $\beta$ is a constant at most of order 1. 
In this case, the two masses $M$ and $M_{\rE}$ are 
related to each other, as 
\be
M=M_{\rE}-6\beta Q^2 M_0\,,
\label{Mre}
\ee
where $M_0 \equiv 4\pi r_s^3\rho_0/3$ corresponds to 
the mass of star with the constant density $\rho_0$. 
The difference between $M$ and $M_{\rE}$ arises 
from the nonvanishing coupling $Q$. 
For $\beta>0$, $M$ is smaller than $M_{\rE}$ 
with the difference $6\beta Q^2 M_0$. 
In the numerical simulation of Fig.~\ref{fig2} ($Q=-1/\sqrt{6}$), 
even when $\rho_0-3P_0$ is negative in the large $\rho_0$ region, 
the constant $\beta$ in Eq.~(\ref{aldef}) is positive outside the star 
and hence $M<M_{\rE}$.
We also compute the mass $M$ from $M_{\rE}$ 
by using the transformation relation (\ref{Mtra}).
As we observe in Fig.~\ref{fig2} (bold dashed line), the mass $M$ 
obtained from the Einstein-frame mass $M_{\rE}$
exactly coincides with the one directly computed in the Jordan frame. 
This shows the consistency of our calculations in 
both Jordan and Einstein frames.

In Fig.~\ref{fig3}, we plot the mass $M$ versus $r_s$ for 
the SLy and FPS EOSs with four different values of $Q$. 
The solid line corresponds to $Q=0$, i.e., the massless 
scalar field in GR. 
As we see in case (b), the modification to $M$ and $r_s$ 
induced by the coupling $Q$ is small for $|Q| \le 0.1$, 
but the difference from the $Q=0$ case arises 
for $|Q|>0.1$. The change of $r_s$ is particularly 
significant for large $|Q|$, like cases (c) and (d) in Fig.~\ref{fig3}.
The mass $M$ is also subject to modifications by the nonvanishing $Q$, but 
the maximum mass $M_{\rm max}$ does not exceed the corresponding 
value in GR for both SLy and FPS EOSs. 
The change of $r_s$ induced by the coupling 
in the range $|Q|~\gtrsim 0.1$ is the main signature of distinguishing between 
massless BD theories and GR from observations.

%%%%%%%%%%%%%%%%%%%%%%%%%%%%%%%%%%%
\section{Brans-Dicke theories with constant scalar mass}
\label{massivesec}
%%%%%%%%%%%%%%%%%%%%%%%%%%%%%%%%%%%

In this section, we study NS solutions in BD theories 
with a constant scalar mass $m$.
This is characterized by the potential
\be
V(\phi)=\frac{1}{2}m^2 \phi^2\,,
\label{qupo}
\ee
in the Jordan frame. 
 
The Starobinsky $f(R)$ model given by 
\be
f(R)=R+\frac{R^2}{6m^2}
\label{fSta}
\ee
falls in this category in the regime $|R| \ll m^2$.
To see this, we use the fact that the scalar degree of freedom  
$\phi$ is related to the Ricci scalar $R$, as Eq.~(\ref{fRre}), 
with $Q=-1/\sqrt{6}$.
In the Starobinsky model, there is the correspondence 
\be
R=3m^2 \left[ e^{\sqrt{6}\phi/(3M_{\rm pl})}-1\right]\,.
\label{Rsta}
\ee
In this case, the scalar potentials in Jordan and Einstein frames 
are given, respectively, by 
\be
V (\phi)=\frac{3}{4}m^2 M_{\rm pl}^2 
\left[ e^{\sqrt{6}\phi/(3M_{\rm pl})}-1\right]^2\,,\qquad
V_{\rE} (\phi)= \frac{3}{4}m^2 M_{\rm pl}^2 
\left[ 1-e^{-\sqrt{6}\phi/(3M_{\rm pl})}\right]^2\,, 
\label{Vsta}
\ee
both of which vanish at $\phi=0$. 
Expanding $V(\phi)$ and $V_{\rE} (\phi)$ around $\phi=0$ 
in the regime $|\phi| \ll M_{\rm pl}$, 
it follows that $V(\phi) \simeq V_{\rE}(\phi) \simeq m^2 \phi^2/2$.
Thus, the scalar field has a constant mass $m$ around the potential minimum. 
In this regime, we have $R \simeq \sqrt{6} m^2 \phi/M_{\rm pl}$ from 
Eq.~(\ref{Rsta}) and hence $|R| \ll m^2$. 
For $|\phi| \lesssim M_{\rm pl}$, the Einstein-Hilbert term $R$ 
dominates over $R^2/(6m^2)$ in the Lagrangian (\ref{fSta}). 
In Refs.~\cite{Cooney:2009rr,Arapoglu:2010rz,Orellana:2013gn,Astashenok:2013vza,Ganguly:2013taa,Yazadjiev,Capozziello:2015yza,Resco:2016upv}, the NS solutions in the model (\ref{fSta}) were studied 
in both Jordan and Einstein frames.

The parameter $a=1/(6m^2)$ in the Starobinsky $f(R)$ model 
is constrained to be $a<5 \times 10^{11}$~m$^2$ from the binary pulsar 
data \cite{Naf:2010zy}. This translates to the bound 
\be
m^{-1} <1.73 \times 10^6~{\rm m}=19.3\,r_0\,.
\label{mbound}
\ee
For general couplings $Q$, the bound on the mass $m$ is 
subject to modifications. As long as $|Q|={\cal O}(0.1)$,  
the order of the upper bound on $m^{-1}$ should be similar 
to that of Eq.~(\ref{mbound}).

The potential in the Einstein frame corresponding to 
Eq.~(\ref{qupo}) is given by 
\be
V_{\rm E} (\phi)=\frac{1}{2}m^2 \phi^2\,
e^{4Q \phi/M_{\rm pl}}\,.
\label{VEpo}
\ee
Outside the star, the field Eq.~(\ref{phiEeq}) obeys
\be
\frac{\rd^2 \phi}{\rd r_{\rE}^2}+\left[ \frac{2}{r_{\rE}}
+\frac{1}{2} \frac{\rd}{\rd r_{\rE}} \ln \left( f_{\rE} h_{\rE} 
\right) \right] 
\frac{\rd \phi}{\rd r_{\rE}}
-\left(1+\frac{2Q \phi}{M_{\rm pl}} \right) 
e^{4Q\phi/M_{\rm pl}}h_{\rE}^{-1}m^2 \phi 
=0\,.
\label{phiEeq2}
\ee
For the potential (\ref{VEpo}), the boundary conditions 
of $\phi$ at spatial infinity correspond to
\be
\phi_{\infty}=0\,,\qquad 
\frac{\rd \phi}{\rd r_{\rE}} (r_{\rE} \to \infty)=0\,.
\label{bouninfE2}
\ee
In the asymptotic regime characterized by $r_{\rE} \gg r_s$, 
we can employ the approximations 
that both $f_{\rm E}$ and $h_{\rm E}$ are close to 1 
with $|Q \phi/M_{\rm pl}| \ll 1$.
Then, the field Eq.~(\ref{phiEeq2}) outside the star 
approximately reduces to
\be
\frac{\rd^2 \phi}{\rd r_{\rE}^2}+\frac{2}{r_{\rE}}
\frac{\rd \phi}{\rd r_{\rE}}-m^2 \phi \simeq 0\,.
\label{phiEeqa}
\ee
This has the following solution 
\be
\phi (r_{\rm E})=c_1 \frac{e^{mr_{\rE}}}{r_{\rE}}
+c_2 \frac{e^{-mr_{\rE}}}{r_{\rE}}\,,
\label{phias}
\ee
where $c_1$ and $c_2$ are integration constants. 
The constant $c_1$ should vanish to satisfy 
the boundary conditions (\ref{bouninfE2}). 
In the exterior region of star close to its surface, 
the solution (\ref{phias}) is subject to modifications. 
Nevertheless, provided that $1+2Q\phi/M_{\rm pl}>0$, 
the last term on the left hand side of Eq.~(\ref{phiEeq2}) 
leads to the growth of $|\phi|$ for the boundary conditions 
satisfying $\phi \neq 0$ or $\rd \phi/\rd r_{\rE} \neq 0$ 
at the surface of star. 
In other words, we require that 
\be
\phi (r_s)=0\,,\quad {\rm and} \quad 
\phi' (r=r_s)=0\,,
\label{phisur}
\ee
to avoid the increase of $|\phi|$ outside the body.  
Unless the boundary conditions (\ref{phisur}) are satisfied, 
the exponential growth of $|\phi|$ starts to occur for the distance 
$r_{\rE} \gtrsim 1/m \equiv r_c$. 
Under the bound (\ref{mbound}), the critical distance $r_c$ 
corresponds to $r_c=19.3 r_0=1.73 \times 10^3$~km, 
which is about $10^2$ times as large 
as the typical radius of NSs ($\sim 10~$km). 
If $1+2Q\phi/M_{\rm pl}<0$ at some distance $r$, then the field 
$\phi$ exhibits damped oscillations with the decreasing amplitude
($|\phi| \propto 1/r_{\rE}$). Then, the system enters the regime 
in which the condition $1+2Q\phi/M_{\rm pl}>0$ is 
satisfied, so the scalar field is eventually subject to 
exponential growth for the distance $r_{\rE} \gtrsim 1/m$. 

The boundary conditions (\ref{phisur}) imply that the field does not 
contribute to the solution outside the body. 
Setting $\phi(r)=0$ and $\phi'(r)=0$  in Eq.~(\ref{Rphi}) for 
the vacuum exterior ($r \geq r_s$), we have $R=0$ and 
hence the external region of star
corresponds to the Schwarzschild geometry. 
This fact was recognized in Ref.~\cite{Ganguly:2013taa} 
for the Starobinsky $f(R)$ model.
{}From Eq.~(\ref{Rphi}), there are the following particular relations 
in $f(R)$ gravity ($Q=-1/\sqrt{6}$):
\be
R=\frac{\sqrt{6}}{M_{\rm pl}} 
e^{-\sqrt{6}\phi/(3M_{\rm pl})}V_{,\phi}\,,\qquad 
R'=\frac{e^{-\sqrt{6}\phi/(3M_{\rm pl})}}
{M_{\rm pl}^2} \left( \sqrt{6} M_{\rm pl} V_{,\phi \phi}
-2V_{,\phi} \right) \phi'\,.
\label{RVp}
\ee
For the quadratic potential (\ref{qupo}) and the Starobinsky potential 
given in Eq.~(\ref{Vsta}), the boundary conditions 
(\ref{phisur}) translate to 
\be
R(r_s=0)=0\,,\quad {\rm and} \quad 
R'(r_s=0)=0\,.
\ee
which coincide with those derived in Ref.~\cite{Ganguly:2013taa} 
by using the junction conditions of $f(R)$ gravity \cite{Deruelle:2007pt}. 

Around the center of star, the scalar field $\phi$ has the radial 
dependence (\ref{phiEinb}) in the Einstein frame. 
In comparison to massless BD theories studied in Sec.~\ref{masslesssec}, 
the potential-dependent term $V_{{\rE},\phi} (\phi_0)$ 
leads to the additional variation of $\phi$ around $r_{\rE}=0$. 
We note that the mass squared $m^2$ 
explicitly appears as one of the coefficients of $r_{\rE}^4$ 
in the expansion of $\phi (r_{\rE})$ in Eq.~(\ref{phiEinb}).
For the boundary conditions where the combination 
$M_{\rm pl}V_{{\rE},\phi}(\phi_0)+Q(\rho_{\rE0}-3P_{\rE0})$ 
is close to 0, the nonvanishing effective mass around the potential minimum 
leads to the variation of $\phi$ with respect to $r$. 
Indeed, this is the case for the chameleon scalar field where 
the variation of $\phi$ occurs mostly around 
the surface of body (``thin shell'') \cite{chame1,chame2,Tsujikawa:2009yf}.

For the massive potential (\ref{qupo}), the field value $\phi_0$ at the center of NSs 
needs to be chosen to satisfy the boundary conditions (\ref{phisur}) at $r=r_s$. 
The EOS of NSs affects the field profile $\phi(r)$ not only around $r=0$ but also 
in the whole interior region of star ($0 \le r \le r_s$).
We numerically solve Eqs.~(\ref{zy})-(\ref{hx}) from the vicinity of $r=0$ 
for both SLy and FPS EOSs with general couplings $Q$. 
For the central matter density in the range $\rho_0>10^{14}$~g\,$\cdot$\,cm$^{-3}$,  
we could not find appropriate values of $\phi_0$ at $r=0$ 
avoiding the exponential growth outside the body.
This means that the boundary conditions (\ref{bouninfE2}) are not 
consistently satisfied at spatial infinity for SLy and FPS EOSs.

Even if we try to fine-tune the field value $\phi_0$ to 
match with the Schwarzschild exterior for some other EOSs, 
it is difficult to realize the boundary conditions of $\phi$ and $\rd \phi/\rd r$ 
which {\it exactly} vanish at $r=r_s$. 
They are highly sensitive to a slight change of the EOS. 
Even when we find a value of $\phi_0$ compatible with the 
conditions (\ref{phisur}) for a particular EOS, the same property 
no longer holds under a tiny change of the EOS. 
Since the EOSs are determined by the nuclear reaction inside NSs, 
a particular EOS chosen to match with the Schwarzschild exterior 
does not generally correspond to the realistic physical EOS \cite{Ganguly:2013taa}.
For the potential (\ref{qupo}) with SLy and FPS EOSs, 
we did not numerically find regular solutions of 
the field and metrics consistent with all the boundary conditions 
at $r=0, r_s, \infty$.

The above discussion is based on the scalar potential (\ref{qupo}), but 
the same conclusion also persists for the Starobinsky $f(R)$ model
given by the Lagrangian (\ref{fSta}). 
In the Starobinsky model we require that $R/(6m^2) \to 0$ 
as $r \to \infty$ for the asymptotic flatness, so the field 
needs to enter the region in which $|\phi/M_{\rm pl}|$ is smaller than 1 
at some distance.
In this regime the scalar potential reduces to $V(\phi) \simeq m^2 \phi^2/2$, 
so the field is eventually subject to exponential growth around the distance 
$r \gtrsim 1/m$ for nonvanishing $\phi(r)$ or $\phi'(r)$ 
outside the star. We performed numerical simulations 
in the Starobinsky model by varying $\phi_0$ at $r=0$ and did not 
find consistent solutions satisfying all the boundary conditions discussed above 
for both SLy and FPS EOSs. The analysis of Ref.~\cite{Ganguly:2013taa} based 
on the polytropic EOS $\rho=\kappa P^{9/5}$ also reached the same conclusion.

In Refs~\cite{Arapoglu:2010rz,Orellana:2013gn,Ganguly:2013taa}, the authors studied 
NS solutions in the $f(R)$ model $f(R)=R-a R^2$, 
where $a$ is a positive constant. In this case, 
the mass squared $m_{\phi}^2$ of the gravitational scalar field
is negative ($m_{\phi}^2 \simeq 1/(3f_{,RR})=-1/(6 a)$). 
Then the field $\phi$ exhibits damped oscillations at large distances, 
so the exponential increase of $|\phi|$ can be avoided at the 
background level. 
However, the linearly perturbed version of Eq.~(\ref{phiEin}) shows that 
the dynamical equation of motion for the field perturbation 
$\delta \phi$ (i.e., the equation associated with the second time 
derivative of $\delta \phi$) contains the negative mass squared, 
which induces the instability of perturbations. 
This property holds not only for the $f(R)$ model $f(R)=R-a R^2$ 
with $a>0$ but also for BD theories with the tachyonic mass squared.

In summary, we showed that BD theories with the positive constant mass squared 
$m^2$ generally face the problem of realizing stable NS field 
configurations consistent with all the boundary conditions. 
Apart from Ref.~\cite{Ganguly:2013taa}, this fact was overlooked in 
most of the past works about NS solutions in $f(R)$ gravity 
with the positive constant $m^2$ outside the star. 
For the massless field studied in Sec.~\ref{masslesssec}, 
there is no need of satisfying the conditions (\ref{phisur}) 
at $r=r_s$ due to the absence of the exponential growing term 
$e^{mr_{\rE}}$ and hence the stable NS solution can be 
easily obtained for a given EOS.

%%%%%%%%%%%%%%%%%%%%%%%%%%%%%%%%%%%
\section{Brans-Dicke theories with self-coupling potential}
\label{selfsec}
%%%%%%%%%%%%%%%%%%%%%%%%%%%%%%%%%%%

For the NS solution discussed in Sec.~\ref{massivesec}, 
the field mass squared $m_{\phi}^2=V_{,\phi \phi}$ is 
a positive constant both inside and outside the star. 
Instead, we can consider other scalar potentials with 
the effective mass depending on the matter density.
This is the case for a chameleon scalar field where the mass is 
large in the region of high density, whereas the field is light 
outside the compact object \cite{chame1,chame2}.
For example, the $f(R)$ models of late-time cosmic 
acceleration \cite{fR1,fR2,fR3,fR4} are designed to have 
a heavy mass in large-curvature regimes to suppress the 
propagation of fifth forces around a compact body 
on the weak gravitational background,  
while the mass of gravitational scalar field is 
as light as today's Hubble expansion rate
$H_0$ \cite{fR1,Faulkner:2006ub,Capozziello:2007eu}. 
In such $f(R)$ models, the existence of relativistic stars was shown 
in Refs.~\cite{Babichev:2009td,Upadhye:2009kt} by considering 
the constant-density profile or polytropic EOS. 

The relativistic star can be also present in BD theories with 
the potential where $V_{,\phi \phi}$ depends on $\phi$. 
Indeed, the numerical simulation of Ref.~\cite{Tsujikawa:2009yf} 
confirmed the existence of relativistic stars with a constant-density profile 
for the inverse power-law potential $V(\phi)=M^{4+n}\phi^{-n}$ ($n>0$).
This is analogous to the chameleon solution on the weak 
gravitational background, in that the scalar field is heavy inside 
the star and its variation occurs mostly 
around its surface (thin shell). 
In the exterior region, the scalar field becomes nearly massless 
due to the significant dropdown of matter density.
In this case, it is possible to avoid the exponential increase of $\phi$  
outside the star by sending $V_{,\phi\phi}$ to 0 as $r \to \infty$.
Thus, we do not need to impose the boundary conditions 
(\ref{phisur}) at $r=r_s$ for the models in which 
the scalar field is nearly massless outside the star.

In what follows, we study NS solutions in BD theories 
with the self-coupling potential 
\be
V(\phi)=\frac{1}{4} \lambda \phi^4\,,
\label{self}
\ee
where $\lambda$ is a positive constant.  
In this case, the second derivative of potential corresponding to 
the mass squared of scalar field is given by 
\be
V_{,\phi\phi}=3 \lambda \phi^2\,,
\label{mp}
\ee
which depends on $\phi$.
The $\phi$ dependence of $V_{,\phi\phi}$ allows 
a possibility for the chameleon mechanism to be at work.
Indeed, this is the case for a compact body 
with the constant density in a weak-field limit \cite{Gubser:2004uf}. 
We extend the analysis to the strong gravitational 
background by using the SLy and FPS EOSs.

If the chameleon mechanism works for a nonrelativistic 
compact body with the constant density $\rho_{\rE}$, 
there is the balance $V_{\rE, \phi}+Q \rho_{\rE}/M_{\rm pl} \simeq 0$ 
in Eq.~(\ref{phiEeq}) except for a thin-shell region near the surface.
On the relativistic background, the pressure $P_{\rE}$ also 
contributes to the field dynamics. Moreover, the realistic NSs have
the $r$-dependent density and pressure. 
At spatial infinity, the scalar field needs to obey the 
boundary conditions (\ref{bouninfE2}). 
In general, the chameleon-like boundary conditions satisfying 
$V_{\rE, \phi}+Q (\rho_{\rE}-3P_{\rE})/M_{\rm pl} \simeq 0$
around the center of NSs do not give rise to the asymptotic 
solutions (\ref{bouninfE2}) on the relativistic background. 
In other words, we need to numerically identify the field 
value $\phi_0$ at $r=0$ consistent with the boundary 
conditions at spatial infinity.

For the potential (\ref{self}), as long as $|\phi|$ gradually approaches 0 
as $r \to \infty$, the second derivative of potential (\ref{mp}) 
also goes to 0. 
This is the way of avoiding the exponential growth of $\phi$ 
outside the body induced by the constant mass term $m$. 
Accordingly, it is not necessary to impose the boundary conditions (\ref{phisur}) 
at the surface of NSs. Unlike the potential $V(\phi)=m^2 \phi^2/2$, 
the absence of the conditions (\ref{phisur}) does not restrict 
the forms of EOSs inside the star.

Using the dimensionless variables introduced in Eq.~(\ref{dimen}), 
the last term ${\cal F}_{\phi} \equiv 
[4QV+V_{,\phi}M_{\rm pl}+Q(\rho-3P)]/(M_{\rm pl}h F)$ 
in Eq.~(\ref{eq3}) can be expressed as
\be
{\cal F}_{\phi}=\frac{M_{\rm pl} e^{2Q \varphi}}
{r_0^2 h} \left[ \tilde{\lambda} \varphi^3 
\left( 1+Q \varphi \right)+8\pi Q 
\left( y-3z \right) \right]\,,
\ee
where 
\be
\tilde{\lambda} \equiv \lambda 
\left( r_0 M_{\rm pl} \right)^2\,,
\ee
with $r_0 M_{\rm pl}=1.107 \times 10^{39}$. 
If $|Q \varphi|$ is smaller than the order 1, the scalar potential modifies 
the field dynamics for $|\tilde{\lambda} \varphi^3| \gtrsim |8\pi Q \left( y-3z \right)|$.
Due to the largeness of the dimensionless quantity $r_0 M_{\rm pl}$, 
even the values of $\lambda$ and $\varphi$ much smaller than 1 can 
give rise to the large contribution to Eq.~(\ref{eq3}). 
For example, we have $|\tilde{\lambda} \varphi^3|>1$ 
for $\lambda^{1/3} \varphi>10^{-26}$.

Numerically, we solve the background equations in the Jordan frame 
by randomly choosing the value of $\phi_0$ at $r=0$ and then identify 
its appropriate value consistent with the boundary conditions 
(\ref{bouninfE2}) by the shooting method. 
For the practical computation, we perform the integration up to the 
distance $r=10^8r_0$ by checking that both $\phi(r)$ and $\phi'(r)$ 
sufficiently approach 0.

%%%%%%%%%%%%%%%%%%%%%%%%%%%%%%%%
\begin{figure}[H]
\begin{center}
\includegraphics[height=3.3in,width=3.4in]{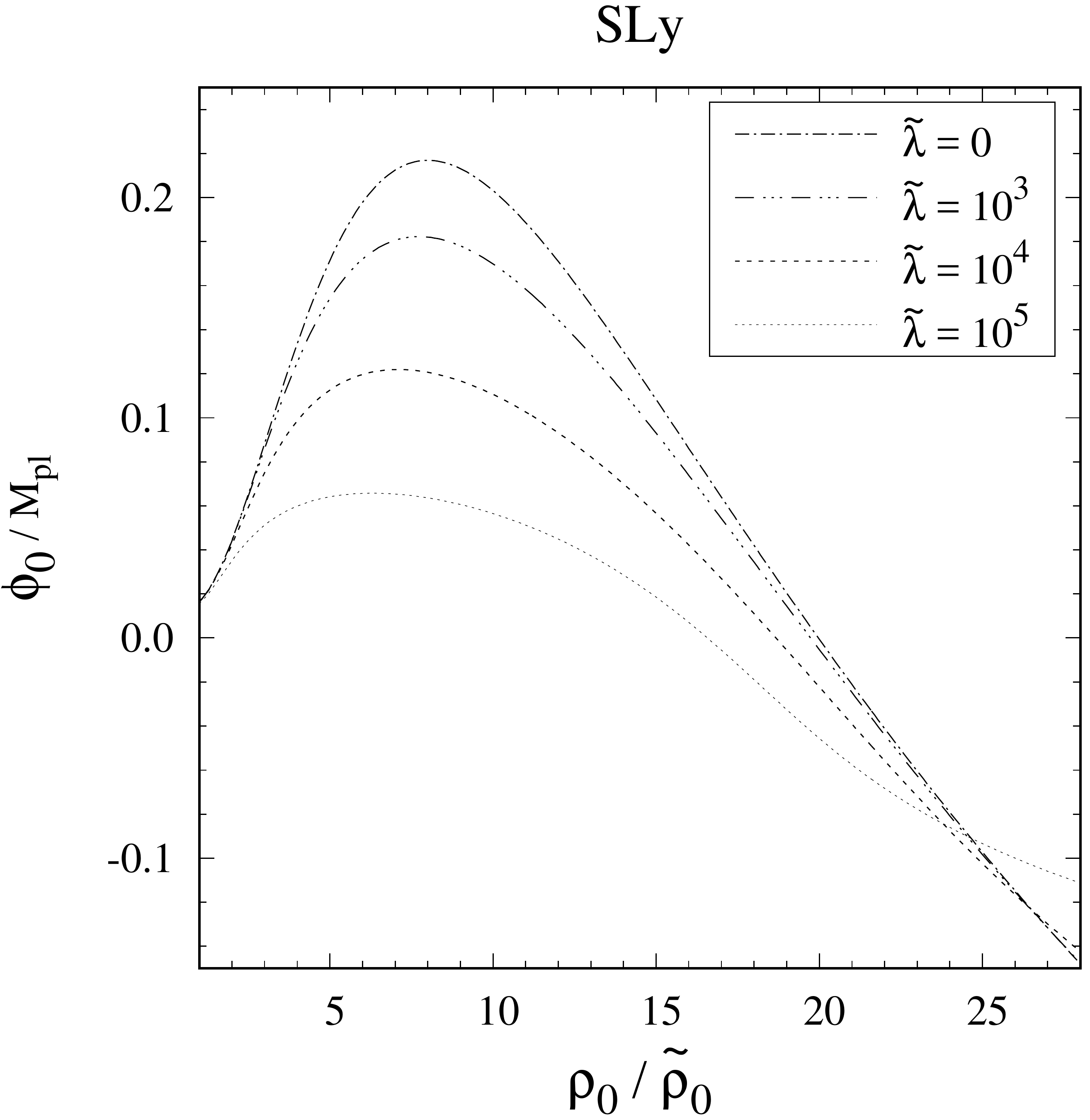}
\includegraphics[height=3.3in,width=3.4in]{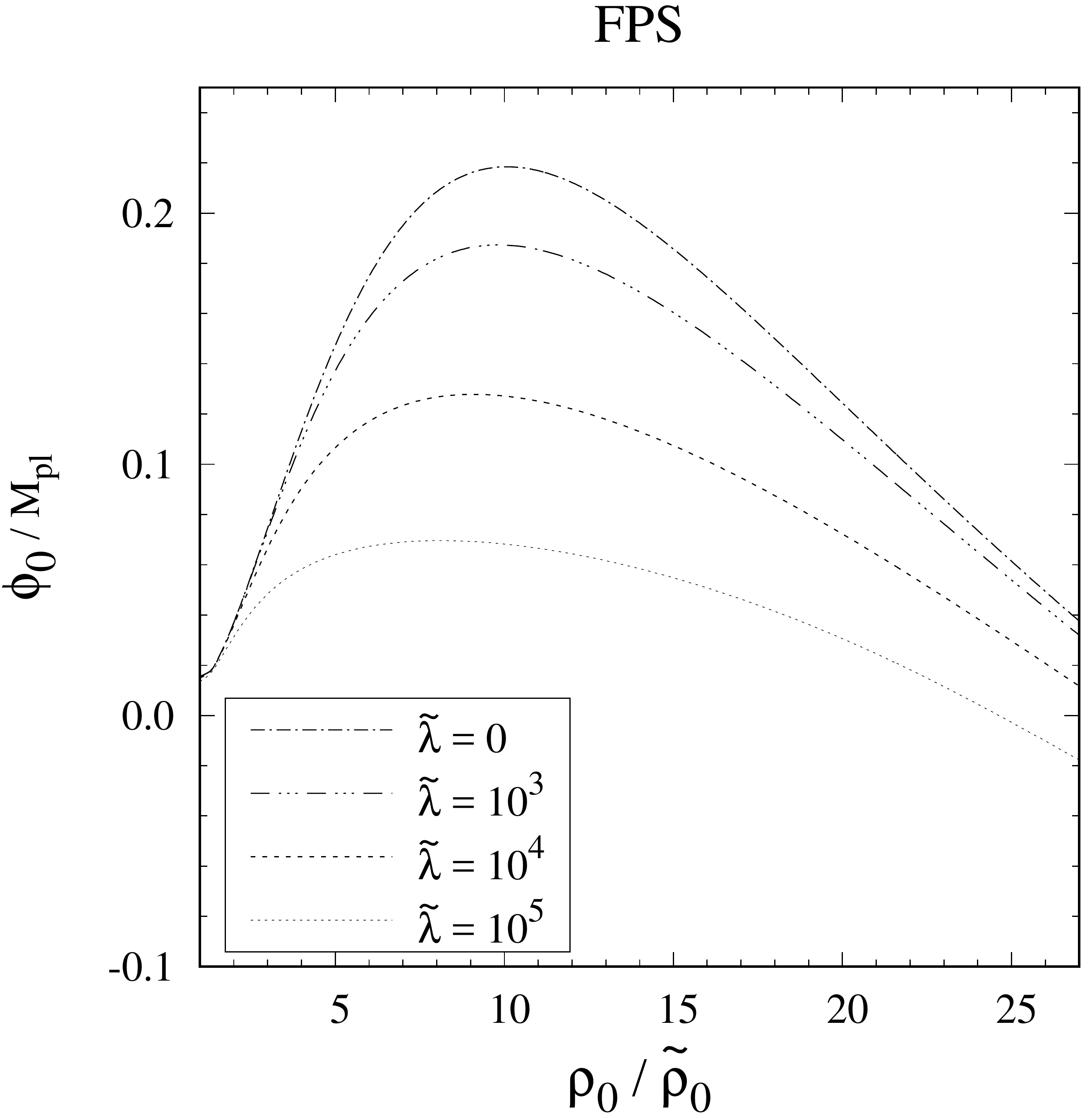}
\end{center}
\caption{\label{fig4} 
The field value $\phi_0$ versus the central density 
$\rho_0$ consistent with the boundary conditions 
(\ref{bouninfE2}) at spatial infinity
for the SLy (left) and FPS (right) EOSs.
Each plot corresponds to BD theories with $Q=-1/\sqrt{6}$ 
and the self-coupling potential (\ref{self}) for four 
different values of $\tilde{\lambda}$.
We choose the boundary conditions (\ref{fexpan})-(\ref{Pexpan}) 
at $s=\ln(r/r_0)=-10$. 
}
\end{figure}
%%%%%%%%%%%%%%%%%%%%%%%%%%%%%%%%

In Fig.~\ref{fig4}, we plot $\phi_0$ versus the central density $\rho_0$ 
for $Q=-1/\sqrt{6}$ with four different values of $\tilde{\lambda}$. 
The EOS is chosen to be SLy (left) and FPS (right). 
The model with $\lambda \neq 0$ can be regarded as $f(R)$ 
gravity with the self-coupling potential (\ref{self}).
The vanishing self-coupling ($\tilde{\lambda}=0$) corresponds to 
the massless scalar field studied in Sec.~\ref{masslesssec}. 
For $\tilde{\lambda} \neq 0$ with a given central density $\rho_0$, 
there exists a unique value of $\phi_0$ consistent with the 
boundary conditions (\ref{bouninfE2})  
for both SLy and FPS EOSs.
When $\rho_0$ is of order $10^{14}$~g\,$\cdot$\,cm$^{-3}$, the condition 
$\rho_{\rE}>3P_{\rE}$ holds and hence the term
$Q(\rho_{\rE}-3P_{\rE})/M_{\rm pl}$ in Eq.~(\ref{phiEeq}) 
is negative for $Q<0$.
Provided that $\phi_0>0$, this matter-coupling term counteracts the increase 
of $\phi$ induced by the potential $V_{{\rm E}, \phi}=3\lambda \phi^2$. 
In the full relativistic region where the opposite inequality  $\rho_{\rE}<3P_{\rE}$ 
holds, the field value $\phi_0$ corresponding to the asymptotic solution 
(\ref{bouninfE2}) is typically negative. 
In Fig.~\ref{fig4}, we observe that, for increasing  $\tilde{\lambda}$, 
there is a tendency that $\phi_0$ decreases (apart from the 
density $\rho_0>4  \times 10^{15}$~g\,$\cdot$\,cm$^{-3}$ for the SLy EOS). 

%%%%%%%%%%%%%%%%%%%%%%%%%%%%%%%%
\begin{figure}[H]
\begin{center}
\includegraphics[height=3.2in,width=3.3in]{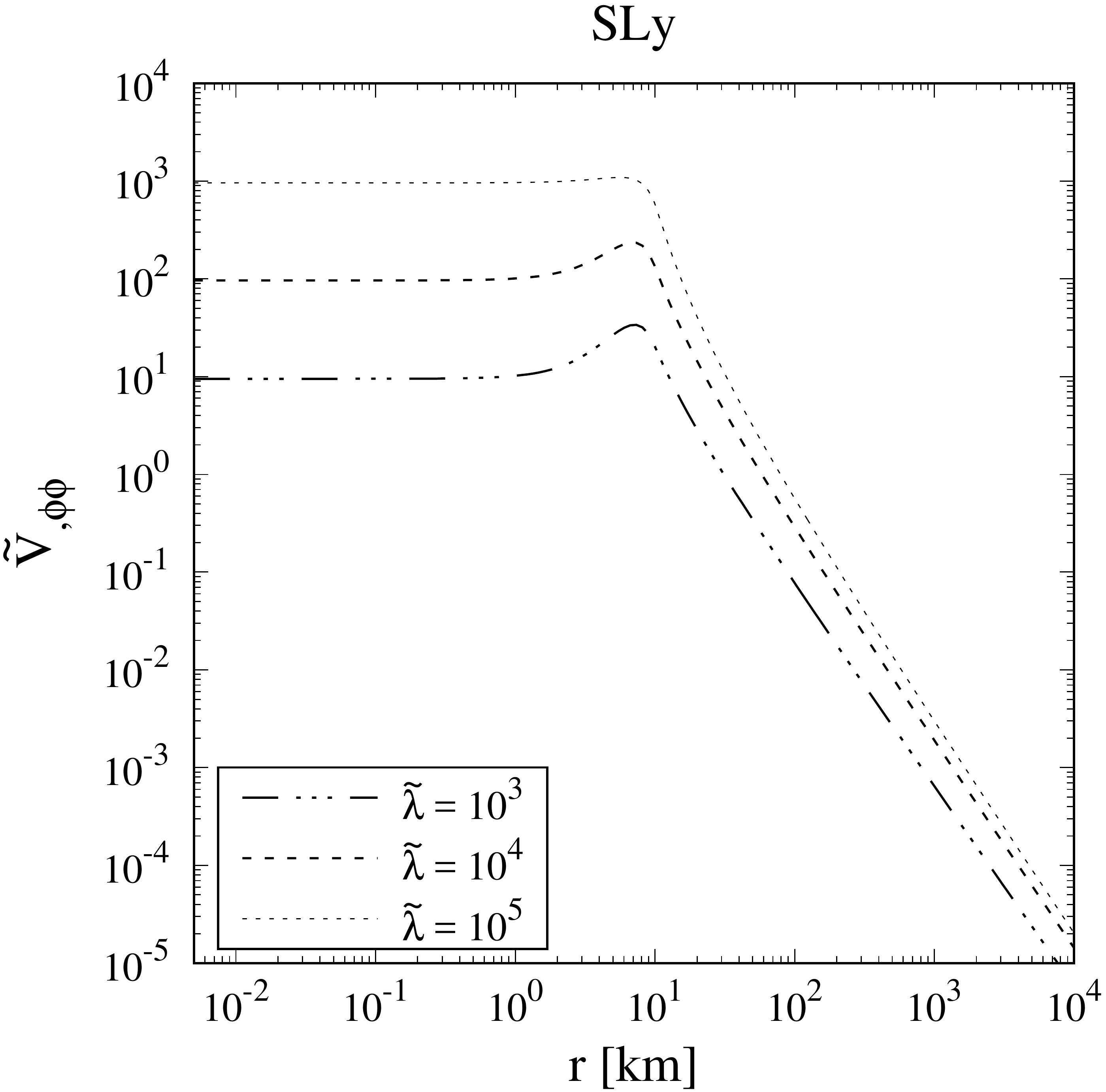}
\includegraphics[height=3.2in,width=3.3in]{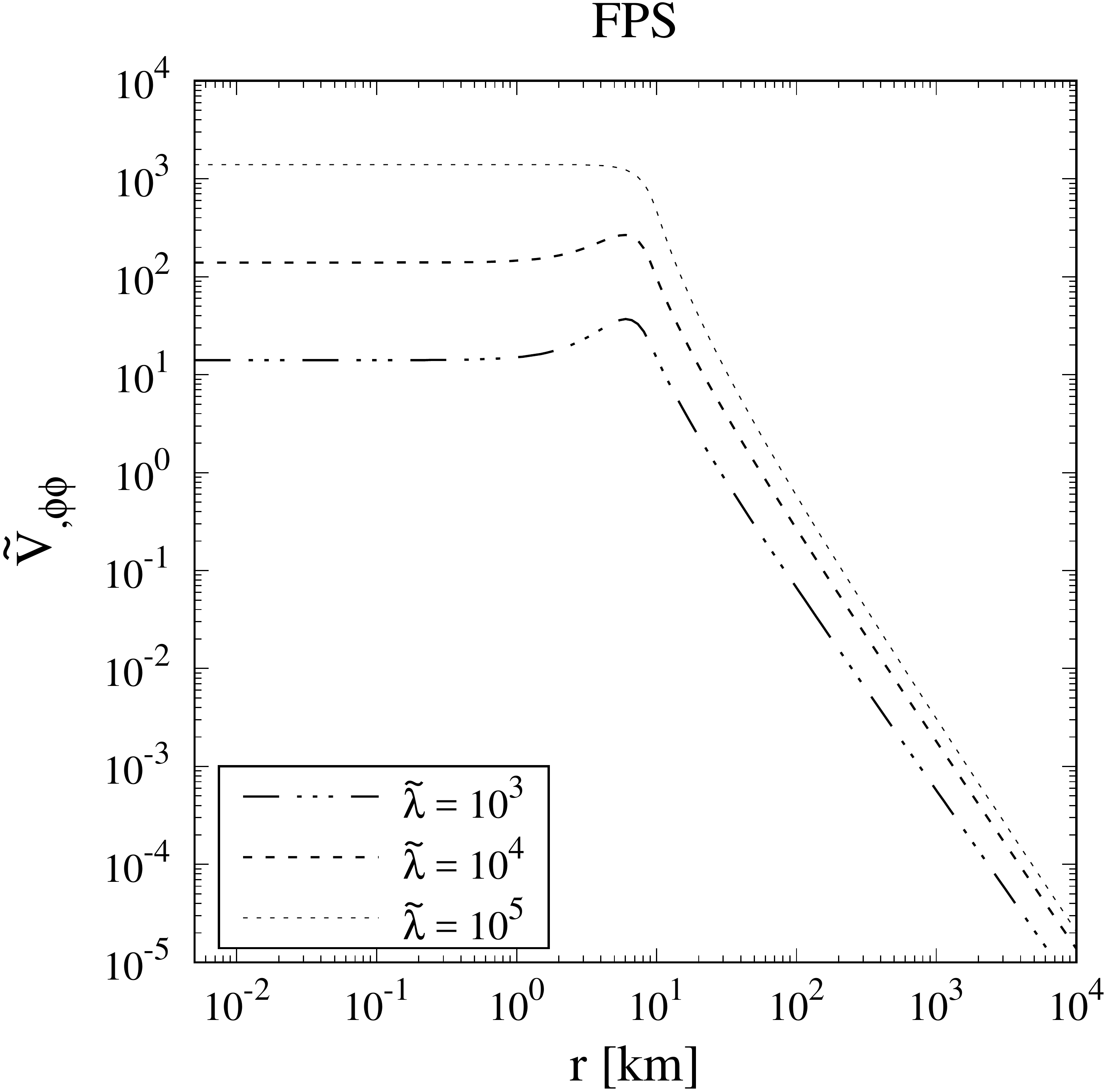}
\end{center}
\caption{\label{fig5}
$\tilde{V}_{,\phi\phi}=V_{,\phi\phi}r_0^2$ 
versus $r$ in BD theories with $Q=-1/\sqrt{6}$ and  $\tilde{\lambda}
=10^3, 10^4, 10^5$ for the SLy (left) and FPS (right) EOSs.
The field value at $r=0$ is chosen to be 
$\phi_0 \simeq 5.6\times10^{-2}\Mpl$ for SLy and 
$\phi_0 \simeq 6.8\times10^{-2}\Mpl$ for FPS. 
}
\end{figure}
%%%%%%%%%%%%%%%%%%%%%%%%%%%%%%%%%

%%%%%%%%%%%%%%%%%%%%%%%%%%%%%%%%
\begin{figure}[H]
\begin{center}
\includegraphics[height=3.4in,width=3.5in]{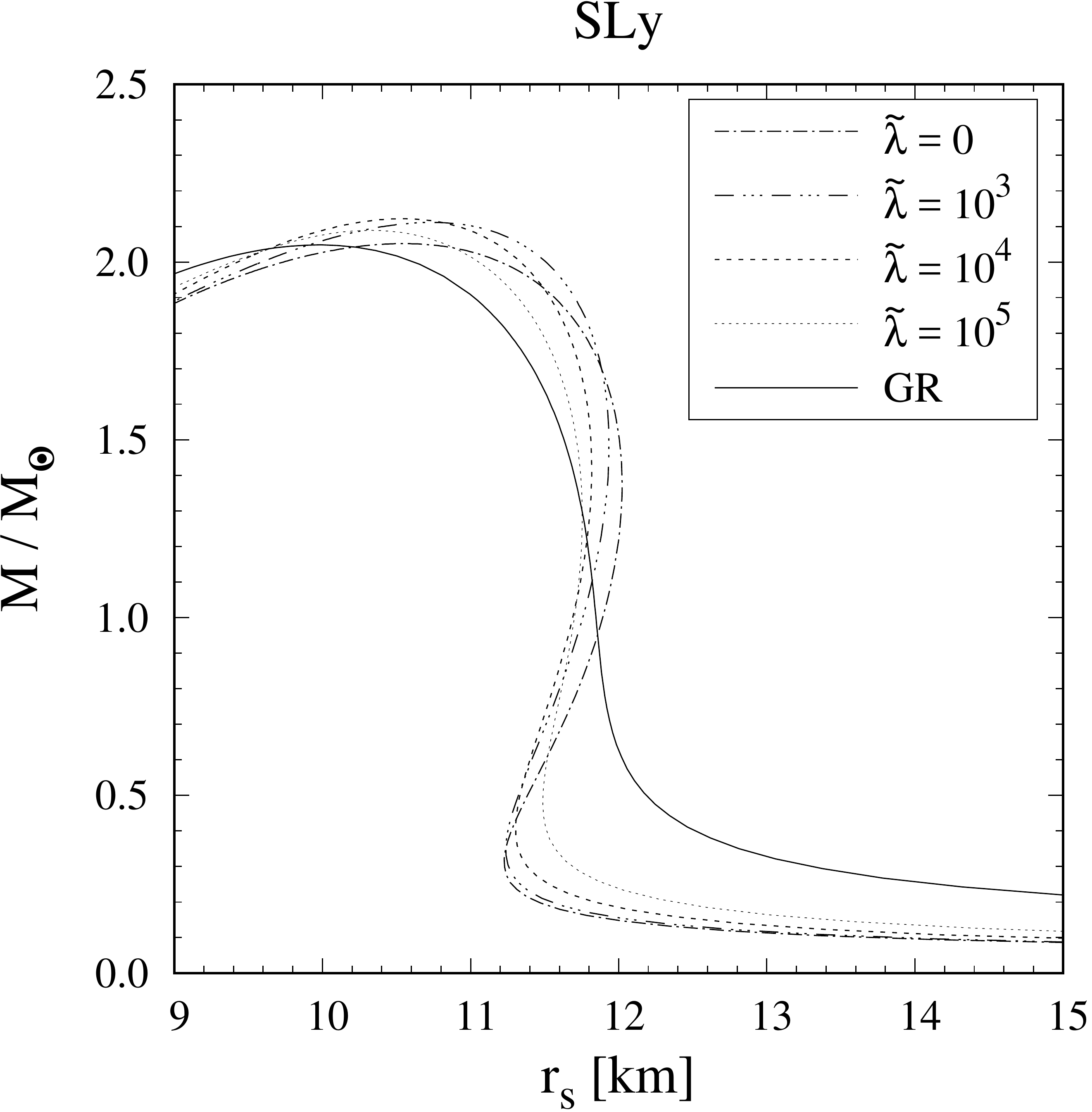}
\includegraphics[height=3.4in,width=3.5in]{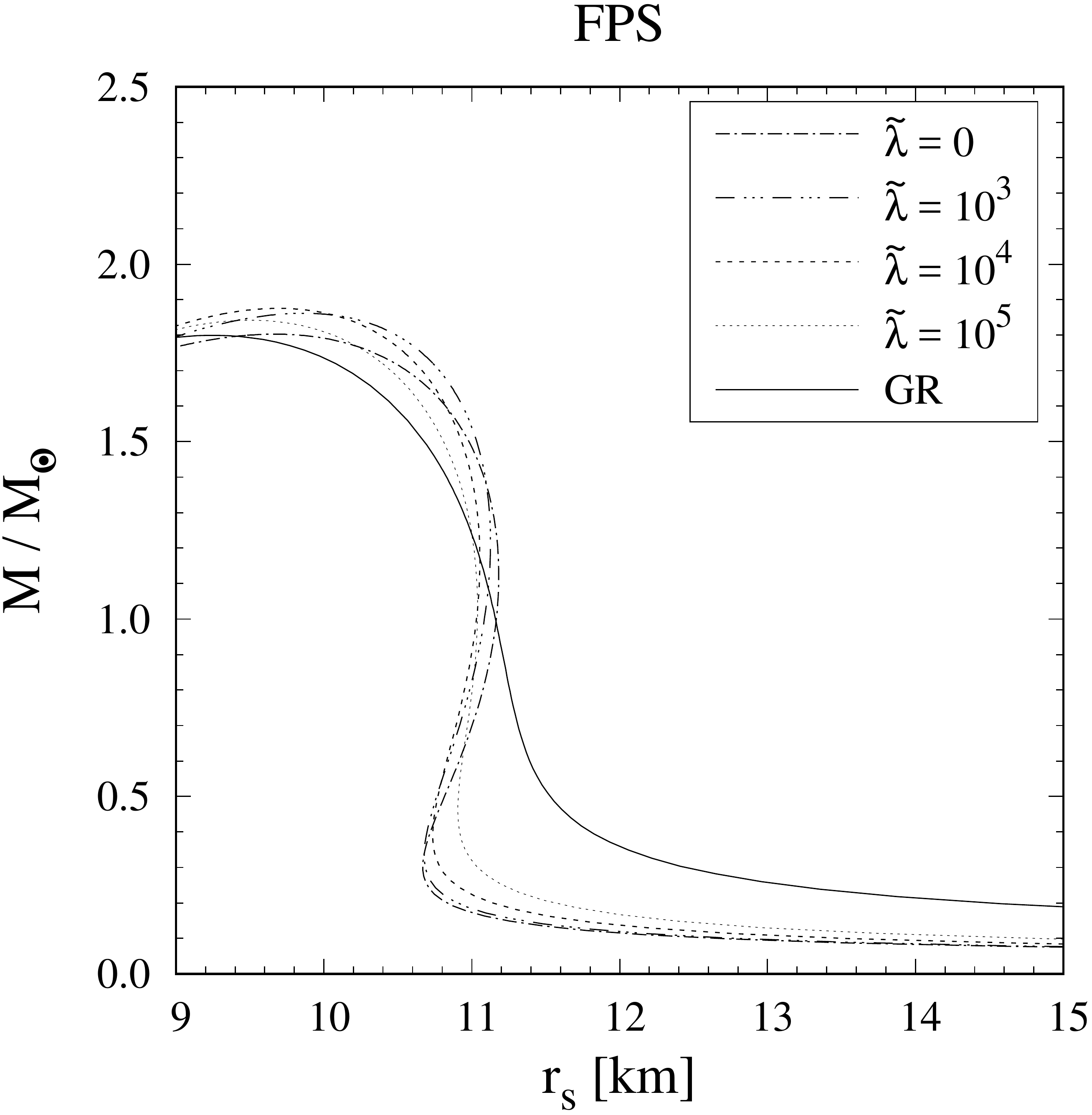}
\end{center}
\caption{\label{fig6}
Mass-radius relation for the SLy (left) and FPS (right) 
EOSs in BD theories with $Q=-1/\sqrt{6}$ and 
the self-coupling potential (\ref{self}). 
Each line corresponds to the relation for four different 
values of $\tilde{\lambda}$ and for GR (solid line).
The boundary conditions are chosen in the same way 
as those in Fig.~\ref{fig4}.
}
\end{figure}
%%%%%%%%%%%%%%%%%%%%%%%%%%%%%%%%%

In Fig.~\ref{fig5}, we show the second derivative of potential 
$\tilde{V}_{,\phi\phi} \equiv V_{,\phi \phi}r_0^2
=3\lambda \phi^2 r_0^2$ versus the distance $r$ 
for several different values of $\tilde{\lambda}$ with $Q=-1\sqrt{6}$.
The field value at $r=0$ is chosen to be 
$\phi_0 \simeq 5.6\times10^{-2}\Mpl$ for SLy and 
$\phi_0 \simeq 6.8\times10^{-2}\Mpl$ for FPS. 
As $\tilde{\lambda}$ increases, $\tilde{V}_{,\phi\phi}$ tends to be larger. 
The variation of $\phi$ inside the NS is not significant except for the 
region around the surface of star. Still, the field profiles in the 
present model are different from that of the chameleon scalar satisfying the relation  
$V_{{\rm E}, \phi}+Q(\rho_{\rE}-3P_{\rE})/M_{\rm pl} \simeq 0$ 
in most internal regions of a compact body. 
Outside the star, $\tilde{V}_{,\phi\phi}$ decreases toward the asymptotic 
value 0. As we observe in Fig.~\ref{fig5}, the field $\phi$ does not vanish 
at the surface of star. Hence the boundary conditions (\ref{phisur}) 
at $r=r_s$ do not hold for BD theories 
with the self-coupling potential. 
In other words, the field $\phi$ contributes to the geometry 
of the external region of star.

In Fig.~\ref{fig6}, we plot the mass-radius relation of NSs 
for $Q=-1/\sqrt{6}$ with four different values of $\tilde{\lambda}$, 
together with the prediction of GR. 
While this is derived by the calculation in the Jordan frame, 
the same mass $M$ and radius $r_s$ can be 
obtained by integrating Eqs.~(\ref{PEeq})-(\ref{MEeq}) 
in the Einstein frame and transforming back to the Jordan frame. 
The case $\tilde{\lambda}=0$ corresponds to 
the massless scalar field plotted in Fig.~\ref{fig2} as the solid lines. 
If $\tilde{\lambda}$ is smaller than the order $10^3$, 
the values of $M$ and $r_s$ are similar to those in the massless case. 
For $\tilde{\lambda} \ge {\cal O}(10^3)$, the difference from the 
$\tilde{\lambda}=0$ case starts to appear for both SLy and FPS EOSs.

As $\tilde{\lambda}$ increases further, the mass-radius relations 
tend to approach that in GR, by reflecting the fact that 
$V_{,\phi\phi}$ gets larger 
inside the star (see Fig.~\ref{fig5}).
For intermediate values of the self-coupling like $\tilde{\lambda}=10^4$, 
the maximum NS mass and the corresponding 
radius are slightly larger than those in GR. 
If we choose a larger coupling $|Q|$ than that in Fig.~\ref{fig6}, 
the modification to $r_s$ tends to be more significant in comparison to 
the change of $M$. This is analogous to the mass-radius relation 
plotted in Fig.~\ref{fig3} for BD theories without the potential.

Finally, we should mention whether the potential (\ref{self}) 
can follow from a specific model of $f(R)$ gravity. 
Let us consider the $f(R)$ Lagrangian 
\be
f(R)=R+aR^p\,,
\label{fRp}
\ee
where $a$ and $p$ are positive constants. 
In this case, the Jordan-frame potential (\ref{fRre}) yields 
$V=M_{\rm pl}^2 a(p-1)R^p/2$ with 
$R=[(e^{\sqrt{6}\phi/(3M_{\rm pl})}-1)/(ap)]^{1/(p-1)}$, i.e., 
\be
V(\phi)=\frac{M_{\rm pl}^2 a(p-1)}{2(ap)^{p/(p-1)}} \left( e^{\sqrt{6}\phi/(3M_{\rm pl})}-1 
\right)^{p/(p-1)}\,.
\ee
In the regime $|\phi| \ll M_{\rm pl}$, this potential approximately reduces to 
\be
V(\phi) \simeq (p-1)V_0 \,\phi^{p/(p-1)}\,,
\label{Vphi0}
\ee
where $V_0$ is a positive constant.
Hence the self-coupling potential (\ref{self}) corresponds to the power 
\be
p=\frac{4}{3}\,.
\ee
For the potential (\ref{Vphi0}), we have 
\be
V_{,\phi\phi}=\frac{pV_0}{p-1} \phi^{\frac{2-p}{p-1}}\,.
\ee
As long as the power $p$ is in the range 
\be
1<p<2\,,
\ee
$V_{,\phi\phi}$ is positive for $\phi>0$. 
Moreover, it has the asymptotic behavior 
$V_{,\phi\phi} \to 0$ as $\phi$ approaches 
0 at spatial infinity. 
Then, as in the self-coupling potential (\ref{self}), the $f(R)$ model (\ref{fRp}) 
with $1<p<2$ can avoid the exponential growth of $\phi$ outside the star. 

In summary, we presented models in BD theories with the potential (\ref{self})
and in $f(R)$ gravity given by the Lagrangian (\ref{fRp}), which can avoid 
the problem of exponential growth of $\phi$ outside the star. 
The existence of regular NS solutions was confirmed in the numerical 
simulation of Fig.~\ref{fig5}. 
As $\lambda$ gets closer to 0, the mass-radius relation tends to deviate 
from that in GR as shown in Fig.~\ref{fig6}. 

%%%%%%%%%%%%%%%%%%%%%%%%%%%%%%%%%%%
\section{Conclusions}
\label{consec}
%%%%%%%%%%%%%%%%%%%%%%%%%%%%%%%%%%%%

In this paper, we studied NS solutions on the spherically symmetric and static 
background in $f(R)$ gravity and BD theories with/without the scalar 
potential $V(\phi)$.
For this purpose, we used the SLy and FPS EOSs given by the parametrization 
(\ref{zeta}) in the Jordan frame. 
In Sec.~\ref{backsec}, we obtained the full background equations in both 
Jordan and Einstein frames together with the solutions of 
metrics, field, and pressure expanded around the center of star. 
The explicit relation between the ADM masses in Jordan and Einstein frames 
is also derived in Eq.~(\ref{Mtra}), which can be used for checking the consistency 
of calculations in two frames.

In Sec.~\ref{masslesssec}, we discussed NS solutions with the scalar field 
profile $\phi(r)$ in BD theories with $V(\phi)=0$. 
As we see in Eq.~(\ref{phiana}), the coupling $Q$ leads to the variation of 
$\phi(r)$ around $r=0$, with the growth $|\phi'(r)| \propto r$. 
The field derivative $|\phi'(r)|$ reaches a maximum value around the surface 
of star ($r=r_s$) and then it starts to decrease for $r>r_s$. 
For the large distance far away from the surface, $|\phi'(r)|$ decreases in proportion 
to $1/r^2$. We found the way of identifying the field value $\phi_0$ at 
$r=0$ consistent with the boundary conditions 
$\phi (r) \to 0$ and $\phi'(r) \to 0$ at spatial infinity. 
We numerically computed the mass-radius relation of NSs 
in the Jordan frame and showed that the calculation in the 
Einstein frame gives the same result after transforming  
back to the Jordan frame. 
The mass-radius relation exhibits the difference from that in GR 
for $|Q|>0.1$. 
As $|Q|$ increases, the radius $r_s$ is subject to large modifications, 
while the maximum mass reached in SLy and FPS EOSs is 
hardly changed in comparison to that in GR.

In Sec.~\ref{massivesec}, we considered BD theories with the quadratic 
potential $V(\phi)=m^2 \phi^2/2$ and studied the effect of mass $m$ 
on the NS configuration.
Far outside the surface of star, the field equation in the Einstein frame 
is of the form (\ref{phiEeqa}), which contains the 
growing-mode solution $e^{m r_{\rE}}/r_{\rE}$. 
To avoid the exponential increase of $|\phi|$ induced by the constant 
mass $m$, the scalar field is restricted to obey 
the boundary conditions (\ref{phisur}) at $r=r_s$. 
This amounts to imposing the Schwarzschild geometry outside the star, without 
the scalar-field contribution to the metric. 
Such boundary conditions are not generally satisfied for arbitrary NS EOSs, 
in which case the field $\phi$ is subject to exponential growth 
for the distance $r \gtrsim 1/m$.
We performed the numerical simulation for both SLy and FPS EOSs
and did not find the NS configuration consistent with all the boundary 
conditions at $r=0, r_s, \infty$. This is also the case for the Starobinsky 
$f(R)$ model given by Eq.~(\ref{fSta}).

In Sec.~\ref{selfsec}, we extended the analysis to BD theories 
with the self-coupling potential $V(\phi)=\lambda \phi^4/4$. 
Since the second derivative of potential $V_{,\phi\phi}=3\lambda \phi^2$
goes to 0 for $\phi(r)$ 
approaching 0 at spatial infinity, it is possible to avoid the 
exponential growth of $\phi$ induced by the mass term 
without imposing the boundary conditions (\ref{phisur})
at $r=r_s$. For given $\lambda$, $Q$, $\rho_0$, 
we identified the field value $\phi_0$ at $r=0$ leading to 
the appropriate boundary conditions $\phi(r) \to 0$ and 
$\phi'(r) \to 0$ as $r \to \infty$.
In general, we found that the chameleon-like boundary 
conditions at $r=0$ do not give rise to the appropriate 
NS solutions for both SLy and FPS EOSs and that, 
even for $\rho<3P$ around $r=0$, there 
are consistent NS configurations. 
We computed the mass-radius relation for $Q=-1/\sqrt{6}$ 
with several different values of $\lambda$ and showed that, 
with increasing $\lambda$, the theoretical curves tend to 
approach that of GR, see Fig.~\ref{fig6}.

We have thus found the new type of NS solutions in BD theories 
in the presence of the self-coupling potential corresponding  
to $f(R)$ theories with the Lagrangian (\ref{fRp}). 
Since the mass-radius relation is different from that in GR, 
the star's compactness parameter $M/r_s$ is subject to modifications.
This leads to the difference for the tidal Love number of 
NSs \cite{Flanagan:2007ix,Damour:2009vw,Binnington:2009bb,Hinderer:2009ca}, 
so that the deviation from GR can be potentially tested in the 
GW observations of NS coalescence like GW170817 \cite{TheLIGOScientific:2017qsa}. 
It is of interest to clarify whether the same property holds in 
BD theories with a wide variety of potentials, e.g., the potential 
accommodating the chameleon mechanism proposed in the context of 
late-time cosmic acceleration \cite{fR1,fR2,fR3,fR4}, by using realistic EOSs, 
and put constraints on theories via tidal Love numbers of NSs.
We leave the detailed analysis for the computation of tidal Love numbers 
in BD theories with the potential and $f(R)$ theories for a future work.

%%%%%%%%%%%%%%%%%
\acknowledgments
%%%%%%%%%%%%%%%%%

RK is supported by the Grant-in-Aid for Young Scientists B 
of the JSPS No.\,17K14297. 
ST is supported by the
Grant-in-Aid for Scientific Research Fund of the JSPS No.\,19K03854 and
MEXT KAKENHI Grant-in-Aid for Scientific Research on Innovative Areas
``Cosmic Acceleration'' (No.\,15H05890).

%%%%%%%%%%%%%%%%%

\end{document}